\documentclass{emulateapj}
\usepackage{natbib}
\usepackage{graphicx}
\usepackage{amssymb}
\usepackage[dvipsnames]{color}

\newcommand{\be}{\begin{equation}}
\newcommand{\ee}{\end{equation}}
\newcommand{\bl}[1]{\mbox{\boldmath$ #1 $}}
\def\ba{\begin{eqnarray}}
\def\ea{\end{eqnarray}}

\def\msun{M_\odot}

\def\ltsima{$\; \buildrel < \over \sim \;$}
\def\simlt{\lower.5ex\hbox{\ltsima}}
\def\gtsima{$\; \buildrel > \over \sim \;$}
\def\simgt{\lower.5ex\hbox{\gtsima}}

\shorttitle{Variable accretion with bursts}
\shortauthors{Vorobyov \& Basu}

\begin{document}

\title{Variable protostellar accretion with episodic bursts}

\author{Eduard~I.~Vorobyov\altaffilmark{1,2} and Shantanu Basu\altaffilmark{3} }

\altaffiltext{1}{Department of Astrophysics, The University of Vienna, Vienna, 1180, Austria; 
eduard.vorobiev@univie.ac.at.} 
\altaffiltext{2}{Research Institute of Physics, Southern Federal University, Stachki 194, 
Rostov-on-Don, 344090, Russia.}
\altaffiltext{3}{Department of Physics and Astronomy, University of Western Ontario,
London, Ontario, N6A 3K7, Canada; basu@uwo.ca.}

\begin{abstract}
We present the latest development of the disk gravitational instability and fragmentation model, 
originally introduced by us to explain episodic accretion bursts in the early 
stages of star formation. Using our numerical hydrodynamics model with improved disk 
thermal balance and star-disk interaction, we computed the evolution of protostellar 
disks formed from the gravitational collapse of prestellar cores. In agreement with
our previous studies, we find that cores of higher initial mass and angular momentum 
produce disks that are more favourable to gravitational instability and fragmentation, 
while a higher background irradiation and magnetic fields moderate the disk tendency to fragment. 
The protostellar accretion in our models is time-variable, thanks to the nonlinear interaction 
between different spiral modes in the gravitationally unstable disk, and can undergo 
episodic bursts when fragments migrate onto the star owing to the gravitational interaction 
with other fragments or spiral arms.  Most bursts occur in the partly embedded Class I phase,
with a smaller fraction taking place in the deeply embedded Class 0 phase and a 
few possible bursts in the optically visible Class II phase. 
The average burst duration and mean luminosity are found to be in good agreement with 
those inferred from observations of 
FU-Orionis-type eruptions. The model predicts the existence of two types of bursts: 
the isolated ones, showing well-defined luminosity peaks separated 
with prolonged periods ($\sim 10^{4}$~yr) of quiescent accretion, and clustered ones, 
demonstrating several bursts occurring one after another during just a few hundred years.  
Finally, we estimate that 40\%--70\% of the star-forming cores can display bursts 
after forming a star-disk system.
\end{abstract}


\section{Introduction}

Low-mass stars form as a result of the gravitational collapse of dense gaseous cores. 
Standard models of core collapse predict that the accretion rate onto 
a forming protostar is proportional to the cube of the sound speed \citep{Larson69,Penston69,Shu77}.
When a finite size of the core is taken into account, numerical simulations 
of gravitationally unstable, spherically symmetric cores indicate that  
accretion is tapering off with 
time in the late evolution \citep{Foster93,VB2005}.
However, when even a modest degree of rotation is present initially in the core 
as suggested by observations \citep[e.g.][]{Caselli2002}, simple arguments based on 
the centrifugal radius and sophisticated numerical hydrodynamics 
simulations both demonstrate that most of the core mass does not fall directly onto the protostar 
but rather lands onto an accretion disk formed from conservation of angular momentum
of the core. 

It has recently become evident that the mass infall rate onto the disk 
$\dot{M}_{\rm infall}$ at radial scales on the order of 1000~AU and the mass accretion rate 
onto the star $\dot{M}$ may be significantly 
different thanks to the complicated interplay of various physical mechanisms of  mass 
and angular momentum transport operating in
the disk.  For instance,  $\dot{M}_{\rm infall}$ in isolated core models gradually 
declines with time from a few $\times 10^{-6}~M_\odot$~yr$^{-1}$ to a negligible value 
by the end of the embedded phase, but $\dot{M}$ in gravitationally unstable disks 
can  be highly variable \citep{Vor2009,Rice2010}, sometimes
exhibiting episodic accretion bursts $\ga 10^{-4}~M_\odot$~yr$^{-1}$ caused by disk gravitational 
fragmentation and migration of the fragments onto the protostar \citep{VB2005,VB2006,VB2010}.  
The formation of giant planets in the disk can also significantly alter the character of accretion,
creating various patterns of variability and bursts exceeding in magnitude 
$10^{-3}~M_\odot$~yr$^{-1}$ \citep{Machida2011,Nayakshin2012}. A combination of the 
magneto-rotational and thermal instabilities in the
inner several AU and gravitational instability further out in the disk (or layered accretion) 
was shown to produce accretion bursts typical for FU-Orionis-type eruptions 
\citep{Armitage2001,Zhu2009}. To complicate the things further, $\dot{M}_{\rm infall}$ may 
experience significant variations when the chaotic and turbulent nature of clustered star formation
is taken into consideration \citep[e.g.][]{Bate2010,Padoan2014}.


Observations support the growing evidence that accretion onto low-mass protostars is at least partly
variable. An ever growing number of FU-Orionis-type and EX-Lupi-like eruptive stars 
\citep{Audard2014} does not fit into the standard models of spherical core collapse. 
The mean/median luminosity of protostars in young star-forming regions appears to be lower 
by about an order of magnitude than that predicted 
from the standard models \citep[e.g.][]{Kenyon1990,Evans2009}. Accretion rates gradually 
declining in time and showing
episodic bursts were shown to resolve this "luminosity problem" \citep{DV2012}. Monitoring of the
accretion variability suggest that about half of all protostars show  up to 50\%  variations in $\dot{M}$
over timescales less than 2~yr \citep{Billot2012}. Surveys of young stars in star-forming regions and
in the Galactic plane revealed that about 0.1\% of objects show a luminosity increase by more than a
factor of 2.5 over 5~yr \citep{Scholz2013}, including recent flares in VSX J205126.1 \citep{Kospal2011}
and V1647 Ori \citep{Abraham2004}. The knotted morphology of jets seen in some
protostellar systems suggests an underlying variability in the mass accretion, 
although the combination of jet velocities and spacing between the knots often suggest shorter 
periods of episodicity than predicted for FU-Orionis-type stars \citep{Arce2013}.

To summarize, accretion onto young stars seems to exhibit a variety of patterns with time variations
of different amplitude and duration and, as noted in \citet{DV2012}, is better termed
as variable accretion with episodic bursts. This newly emerging paradigm is beginning to supersede 
the classical Shu-Larson-Penston steady accretion models and may have important consequences for 
the evolution of stars and planets. For instance, variable accretion can help to explain 
the luminosity spread of young clusters without invoking a significant age spread 
\citep{Baraffe2009,Baraffe2012}. In addition, quiescent periods between accretion 
bursts can promote 
disk fragmentation and giant planet formation \citep{Stamatellos2011}. Finally, variable accretion
with episodic bursts is expected to have a significant impact on the disk and envelope chemistry
and on the composition of ices in protostellar disks \citep{Lee2007,Visser2012,Kim2012}

In this paper, we revisit the disk instability and fragmentation model for episodic accretion and luminosity
bursts, originally developed by us in a series of papers \citep{VB2005,VB2006,VB2010}, using an 
improved numerical hydrodynamics code which takes into account a better disk thermal physics, improved
dust opacities, and  an accurate calculation of the stellar photospheric and accretion 
luminosities using a stellar evolution code that takes stellar accretion into account. 
The latter update allows us to
calculate the burst statistics and perform direct comparison with observations, 
and also make prediction regarding the expected fraction of star-forming cores that can display 
bursts after forming a star-disk system.
The paper is organized as follows. A brief description of the numerical model and recent updates
are presented in Section~\ref{model}. The main results are described in Section~\ref{results}.
The characteristics of the bursts obtained in the framework of our model are reviewed
in Section~\ref{bursts}. The time evolution of individual bursts is considered in Section~\ref{cluster}.
The expected fraction of star-forming cores than can exhibit bursts after forming a star-disk
systems is calculated in Section~\ref{fraction} and main conclusions are summarized in Section~\ref{summary}.


\section{Model description}
\label{model}
Our numerical model is described in detail in 
\citet{VB2010} and is briefly reviewed below with the emphasis on several recent updates.
We start our numerical simulations from the gravitational collapse of a {\it starless} cloud core,
continue into the embedded phase of star formation, during which
a star, disk, and envelope are formed, and terminate our simulations when the age of the
star becomes older than 1.0~Myr. Such long integration times are made possible by the use
of the thin-disk approximation, the justification of which is provided in \citet{VB2010}.
The protostellar disk occupies the inner part of the numerical polar grid
and is exposed to intense mass loading from the infalling envelope.  

To avoid too small time steps, we introduce a ``sink cell'' at $r_{\rm sc}=6.0$~AU and 
impose a free inflow inner boundary condition
and a free outflow outer boundary condition so that the matter is allowed to flow out of 
the computational domain but is prevented from flowing in. 
The sink cell is dynamically inactive; it contributes only to the total gravitational 
potential and secures a smooth behaviour of the gravity force down to the stellar surface.
During the early stages of the core collapse, we monitor the gas surface density in 
the sink cell and when its value exceeds a critical value for the transition from 
isothermal to adiabatic evolution, we introduce a central point-mass object.
In the subsequent evolution, 90\% of the gas that crosses the inner boundary 
is assumed to land on the central object. 
The other 10\% of the accreted gas is assumed to be carried away with protostellar jets. 

\subsection{Main equations}

The basic equations of mass, momentum, and energy transport in the thin-disk limit are
\begin{equation}
\label{cont}
\frac{{\partial \Sigma }}{{\partial t}} =  - \nabla_p  \cdot 
\left( \Sigma \bl{v}_p \right),  
\end{equation}
\begin{eqnarray}
\label{mom}
\frac{\partial}{\partial t} \left( \Sigma \bl{v}_p \right) &+& \left[ \nabla \cdot \left( \Sigma \bl{v_p}
\otimes \bl{v}_p \right) \right]_p =   - \nabla_p {\cal P}  + \Sigma \, \bl{g}_p + \\ \nonumber
& + & (\nabla \cdot \mathbf{\Pi})_p, - \nabla_p \left({B_z^2 \over 4 \pi}\, Z \right)
 +  {B_z {\bl B}_p^+ \over 2 \pi}
\label{energ}
\end{eqnarray}
\begin{equation}
\frac{\partial e}{\partial t} +\nabla_p \cdot \left( e \bl{v}_p \right) = -{\cal P} 
(\nabla_p \cdot \bl{v}_{p}) -\Lambda +\Gamma + 
\left(\nabla \bl{v}\right)_{pp^\prime}:\Pi_{pp^\prime}, 
\end{equation}
where subscripts $p$ and $p^\prime$ refers to the planar components $(r,\phi)$ 
in polar coordinates, $\Sigma$ is the mass surface density, $e$ is the internal energy per 
surface area, 
${\cal P}$ is the vertically integrated gas pressure calculated via the ideal equation of state 
as ${\cal P}=(\gamma-1) e$,
$Z$ is the radially and azimuthally varying vertical scale height
determined in each computational cell using an assumption of local hydrostatic equilibrium,
$\bl{v}_{p}=v_r \hat{\bl r}+ v_\phi \hat{\bl \phi}$ is the velocity in the
disk plane, and $\nabla_p=\hat{\bl r} \partial / \partial r + \hat{\bl \phi} r^{-1} 
\partial / \partial \phi $ is the gradient along the planar coordinates of the disk. 
The gravitational acceleration in the disk plane, $\bl{g}_{p}=g_r \hat{\bl r} +g_\phi \hat{\bl \phi}$, takes into account self-gravity of the disk, found by solving for the Poisson integral 
\citep[see details in][]{VB2010}, and the gravity of the central protostar when formed. 
Turbulent viscosity is taken into account via the viscous stress tensor 
$\mathbf{\Pi}$, the expression for which is provided in \citet{VB2010}.
We parameterize the magnitude of kinematic viscosity $\nu$ using the $\alpha$-prescription 
with a spatially and temporally uniform $\alpha$. 

Two l.h.s. terms in Equation~(\ref{mom}) represent magnetic pressure and tension 
in the thin-disk approximation, where $B_z$ is the vertically uniform magnetic field in the disk
and  ${\bl B}_p^+=B_r^+ \hat{\bl r}+ B_\phi^+ \hat{\phi}$ are the planar components of the magnetic field at the top surface of the disk. 
In the flux-freezing approximation adopted in this work
the vertical magnetic field component in the disk can be determined from the relation
$B_z=2\pi G^{1/2} \Sigma/ \mu_{\rm B}$ \citep{Nakano78}, where $\mu_{\rm B}$ is the spatially 
uniform mass-to-flux ratio.
The planar components of the magnetic field  
are directly related to the planar components of gravitational acceleration ${\bl g}_p$ 
through the following relation $B_z {\bl B}_p^+/(2 \pi)=-{\bl g}_p/\mu_{\rm B}^2$
 \citep[see][for more details]{VB2006}.


The radiative cooling $\Lambda$ in equation~(\ref{energ}) is determined using the diffusion
approximation of the vertical radiation transport in a one-zone model of the vertical disk 
structure \citep{Johnson03}
\begin{equation}
\Lambda={\cal F}_{\rm c}\sigma\, T_{\rm mp}^4 \frac{\tau}{1+\tau^2},
\end{equation}
where $\tau$ is the optical depth to the disk midplane, $\sigma$ is the Stefan-Boltzmann constant, 
$T_{\rm mp}={\cal P} \mu / R \Sigma$ is 
the midplane temperature of gas\footnote{This definition of the midplane temperature is accurate within
a factor of unity \citep{Zhu2012}}, $\mu=2.33$ is the mean molecular weight, $R$ is the universal 
gas constant, and ${\cal F}_{\rm c}=2+20\tan^{-1}(\tau)/(3\pi)$ is a function that 
secures a correct transition between the optically thick and optically thin regimes. 
The heating function is expressed as
\begin{equation}
\Gamma={\cal F}_{\rm c}\sigma\, T_{\rm irr}^4 \frac{\tau}{1+\tau^2},
\end{equation}
where $T_{\rm irr}$ is the irradiation temperature at the disk surface 
determined by the stellar and background black-body irradiation as
\begin{equation}
T_{\rm irr}^4=T_{\rm bg}^4+\frac{F_{\rm irr}(r)}{\sigma},
\label{fluxCS}
\end{equation}
where $T_{\rm bg}$ is the uniform background temperature (in our model set to the 
initial temperature of the natal cloud core)
and $F_{\rm irr}(r)$ is the radiation flux (energy per unit time per unit surface area) 
absorbed by the disk surface at radial distance 
$r$ from the central star. The latter quantity is calculated as 
\begin{equation}
F_{\rm irr}(r)= \frac{L_\ast}{4\pi r^2} \cos{\gamma_{\rm irr}},
\label{fluxF}
\end{equation}
where $\gamma_{\rm irr}$ is the incidence angle of 
radiation arriving at the disk surface (with respect to the normal) at radial distance $r$.
The stellar luminosity $L_\ast$ is the sum of the accretion luminosity 
$L_{\rm \ast,accr}=(1-\epsilon) G M_\ast \dot{M}/2
R_\ast$ arising from the gravitational energy of accreted gas and
the photospheric luminosity $L_{\rm \ast,ph}$ due to gravitational compression and deuterium burning
in the stellar interior. The stellar mass $M_\ast$ and accretion rate onto the star $\dot{M}$
are determined using the amount of gas passing through
the sink cell, while the stellar radius $R_\ast$ is returned
by a stellar evolution code (see Section~\ref{updates} for details and 
the definition of $\epsilon$). Equations~(\ref{cont})--(\ref{energ}) are solved 
in the polar coordinates on a numerical grid  with $512 \times 512$ grid zones. 
The solution procedure is described in detail in \citet{VB2010}.

\begin{table*}
\renewcommand{\arraystretch}{1.2}
\center
\caption{Model parameters}
\label{table1}
\begin{tabular}{cccccccccc }
\hline\hline
Model & $M_{\rm core}$ & $\beta$ & $T_{\rm init}$ & $\Omega_0$  & $r_{\rm 0}$ & $\Sigma_0$  
& $R_{\rm out}$ & $\alpha$ & $\mu_{\rm B}$ \\
 & ($M_\odot$) & ($\%$) & (K) & (km~s$^{-1}$~pc$^{-1}$) &  (AU) & (g~cm$^{-2}$) & (pc) & & \\
\hline
1 & 1.1 & 0.88 & 10 & 1.4  & 2400 & $5.2\times 10^{-2}$ & 0.07 & $5\times10^{-3}$ & 0 \\
2 & 1.5 & 0.88 & 10 & 1.0  & 3400 & $3.7\times 10^{-2}$ & 0.1 & $5\times10^{-3}$ & 0 \\
3 & 0.31 & 0.88 & 10 & 3.3  & 685 & $1.8\times 10^{-1}$ & 0.03 & $5\times10^{-3}$ & 0 \\
4 & 1.1 & 0.14 & 10 &  0.57  & 2400 & $5.2\times 10^{-2}$ & 0.07 & $5\times10^{-3}$ & 0 \\
5 & 1.1 & 0.88 & 25 & 5.7  & 960 & $3.2\times 10^{-1}$ & 0.028 & $5\times10^{-3}$ & 0 \\
6 & 1.1 & 0.88 & 10 & 1.4  & 2400 & $5.2\times 10^{-2}$ & 0.07 & $3\times10^{-2}$ & 0 \\
7 & 1.1 & 0.88 & 10 & 1.4  & 2400 & $5.2\times 10^{-2}$ & 0.07 & $5\times10^{-3}$ & 3.33\\
\hline
\end{tabular}
\end{table*}

\subsection{Recent updates}
\label{updates}
In this study, several important updates have been implemented to the numerical code 
as compared to the earlier work of \citet{VB2010}. First, we have implemented newer  
Semenov dust opacities \citep{Semenov2003} instead of older Bell \& Lin opacities \citep{Bell94}.
The Semenov opacities are somewhat higher than those of Bell \& Lin in the temperature range typical
for protostellar disks, which results in a somewhat higher gas temperature. 
Second, we considered a stiffer equation of state taking into account the
fact that the rotational and vibrational degrees of freedom of molecular hydrogen are excited
only above 100~K \citep[e.g.][]{MI2000}. As a result, the ratio of specific heats takes the following
from
\begin{equation}
\gamma  = \left\{ \begin{array}{ll} 
   5/3,  &\,\,\,  \mbox{if $T_{\rm g}<100$~K }, \\ 
   7/5,  & \,\,\, \mbox{if $100~\mathrm{K} \le T_{\rm g}<2000$~K} ,  \\
   1.1,  & \,\,\, \mbox{if $T_{\rm g}\ge2000$~K}.
   \end{array} 
   \right. 
   \label{function} 
\end{equation}
In \citet{VB2010}, $\gamma$ was set to 7/5 for $T_{\rm g}<2000$~K.  The net result is
an overall moderate increase in the disk temperature at distances $\ga 10$~AU.

The aforementioned updates enable a better calculation of the thermal balance in the disk
and hence a more accurate study of the gravitational instability and fragmentation.
We note that the net effect of these updates is an increase in the disk temperature, 
making disk fragmentation more difficult.

Finally, we have improved on the method in which the parameters of the central star, 
are calculated. 
In \citet{VB2010}, pre-main-sequence tracks for
the {\it non-accreting} low-mass stars and brown dwarfs  of \citet{Dantona94} were employed
to calculate the stellar photospheric luminosity and radius.
In this work, the properties of the forming protostar are 
calculated using a stellar evolution code described in \citet{Baraffe2010}.
As in \citet{Baraffe2012}, we assume that the fraction $\epsilon$ of the accretion energy  
$G M_\ast{\dot M}/(2 {R_\ast})$ is absorbed by the protostar, while the fraction 
$(1-\epsilon$) is radiated away and contributes to the accretion luminosity of the star
$L_{\rm \ast,accr}$. 
Despite many efforts, the exact value of $\epsilon$ in low-mass star formation is not 
known. In the present calculations, we adopt a so-called "hybrid" scheme 
\citep[see][for detail]{Baraffe2012} with $\epsilon=0$ when accretion rates remain 
smaller than  a critical value $\dot{M}_{\rm cr}=10^{-5}~M_\odot$~yr$^{-1}$, and 
$\epsilon=0.2$  when $\dot{M} >  \dot{M}_{\rm cr}$.

The stellar evolution  code is coupled with the main hydrodynamical code in real time. 
The input parameter to the stellar evolution code provided by disk modeling is 
the mass accretion rate onto the star $\dot{M}$.
The output of the stellar evolution code are the stellar radius $R_\ast$ and the photospheric 
luminosity $L_{\ast,\rm ph}$, which are employed by the disk hydrodynamics simulations 
to calculate the total stellar luminosity and the radiation flux reaching the disk surface.
Due to heavy computational load  the stellar evolution code is 
invoked to update the properties of the protostar 
only every 20~yr, while the hydrodynamical time step may be as small
as a few weeks and the entire duration of numerical simulations may exceed 1.0 Myr.

The coupling of disk modeling with the stellar evolution code is essential. 
As was demonstrated in \citet{Baraffe2009,
Baraffe2012}, the stellar properties derived from evolution
models that do not take stellar accretion into account (such as \citet{Dantona94} data 
used in our previous work) can significantly differ from those
derived from accreting models. For instance, 
the stellar photospheric luminosities may differ by a factor of up to 10 (see
fig.~5 in Baraffe et al. 2012) and stellar radii\footnote{Stellar radius affects 
the proper calculation of the accretion luminosity.} by a factor of several (see fig.~2
in Baraffe et al. 2009), depending on the stellar mass and the fraction of accretion 
energy absorbed by the star $\epsilon$.  Therefore, employing accreting stellar evolution models
is crucial for an accurate comparison of burst characteristics derived from numerical modeling 
with those 
measured in young star-forming regions (Sections \ref{bursts}-\ref{fraction}). 
In addition, they can also enable a better calculation of the disk thermal physics,
because the main source of heating for flared disks at 
distances where fragmentation takes place is mostly stellar 
irradiation (viscosity is more important in the inner several tens of AU).


More details on the coupling of the two numerical codes (the hydrodynamic and 
stellar evolution ones) can be found in \citet{Vorobyov2013}.


\subsection{Initial conditions} 

For the initial distribution of the gas surface density 
$\Sigma$ and angular velocity $\Omega$, we adopted those derived by \citet{Basu1997} for 
pre-stellar cores formed as a result of ambipolar diffusion, with the angular
momentum remaining constant during axially-symmetric core compression
\begin{equation}
\Sigma={r_0 \Sigma_0 \over \sqrt{r^2+r_0^2}}\:,
\label{dens}
\end{equation}
\begin{equation}
\Omega=2\Omega_0 \left( {r_0\over r}\right)^2 \left[\sqrt{1+\left({r\over r_0}\right)^2
} -1\right].
\label{omega}
\end{equation}
Here, $\Omega_0$ and $\Sigma_0$ are the angular velocity and gas surface
density at the center of the core and $r_0 =\sqrt{A} c_{\rm s}^2/\pi G \Sigma_0 $
is the radius of the central plateau, where $c_{\rm s}$ is the initial sound speed in the core. 
The gas surface density distribution described by equation~(\ref{dens}) can
be obtained (to within a factor of unity) by integrating the 
three-dimensional gas density distribution characteristic of 
Bonnor-Ebert spheres with a positive density-perturbation amplitude A \citep{Dapp09}.
In all models the value of the initial density enhancement $A$ is set to 1.2
and all cores have a fixed ratio $r_{\rm out}/r_0=6.0$, where $r_{\rm out}$ is the 
radius of the core, implying that the cores are initially unstable to gravitational collapse.

Individual models are generated by first choosing $r_{\rm out}$ and then calculating
$\Sigma_0$. The central angular velocity $\Omega_0$ is chosen so as to
generate cores with different ratios of rotational to gravitational energy $\beta$,
consistent with the values inferred for pre-stellar cores by \citet{Caselli2002}.
We have considered 7 models, the parameters of which are listed in Table~\ref{table1}.
The prototype model~1 has the core mass $M_{\rm core}$ set to 1.1~$M_\odot$, 
the value of $\beta$ set to 0.88\%, and the viscous $\alpha$-parameter set to $5\times10^{-3}$. 
The initial gas temperature is $T_{\rm init}=10$~K and magnetic fields are turned off.
Other models are compared against model~1, with their parameters being varied so as 
to emphasize the effect of different initial core masses, temperatures, 
and rotation rates, as well as the effect of turbulent viscosity and magnetic field. 


\section{Variable accretion with episodic bursts} 
\label{results}
We start with comparing the long-term evolution of circumstellar disks in our non-magnetic models,
continue with analysing the accretion rates and the burst statistics in different models, 
and finish with considering the effect of magnetic fields.

\subsection{Disk evolution}
\label{disks}
\begin{figure*}
 \centering
  \includegraphics[width=13.5cm]{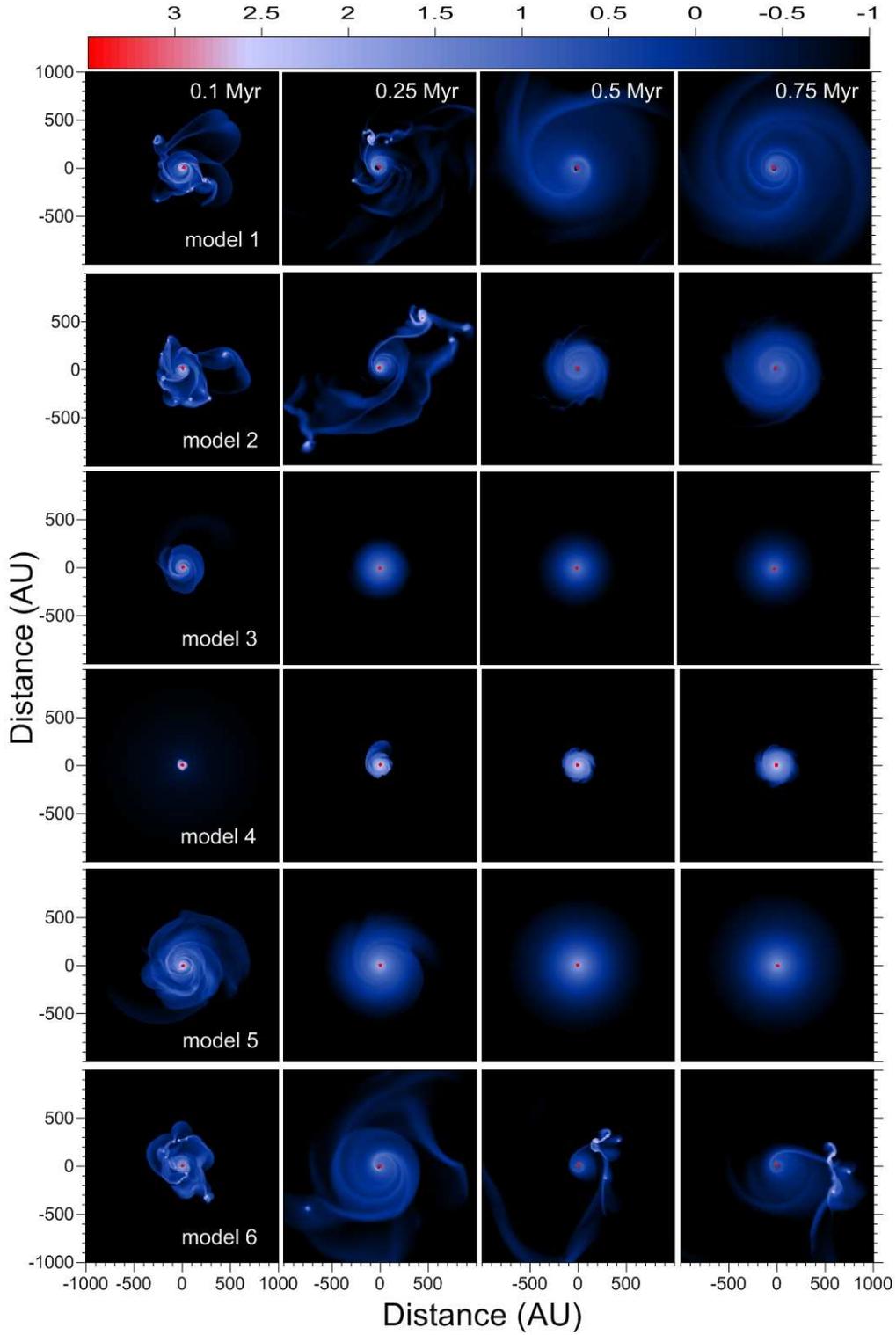}
  \caption{Images of the gas surface density in the inner $2000\times2000$~AU box in models~1--7 (from
  top to bottom). Each row represents an individual model at four characteristic times after the
  formation of the central star. The star is marked by the red circle in the coordinate center. 
  The scale  bar is in log g~cm$^{-2}$}
  \label{fig1}
\end{figure*}

Figure~\ref{fig1} presents the time evolution of the  gas surface density $\Sigma$ 
in seven models, each row of images corresponding to a particular model.
Individual images show the inner region of $2000$~AU on each side, whereas 
the total computational domain is usually much greater. 
The minimum value of the density scale is set to -1.5 
(in log units of g~cm$^{-2}$), a typical value at the disk outer edge \citep{Vor2010}
Four columns show the disk evolution at four representative 
times $t$ elapsed since the formation of the central protostar.

We start by analyzing the prototype model~1. Evidently, the early disk evolution ($t$=0.15--0.25~Myr)
in this model is characterized by vigourous gravitational instability and fragmentation. The disk has
an irregular spiral structure with multiple fragments forming in the densest parts of the arms.
As time progresses to $t=$0.5--0.75~Myr, the disk becomes less irregular, taking a smoother 
shape and exhibiting only a week spiral structure. No fragments are visible in the disk at that time,
which does not necessarily mean that disk fragmentation has ceased completely. 
As will be shown below, this is merely an artifact of infrequent time sampling in Figure~\ref{fig1}
and short-lived fragmentation episodes continue to occur even in the late disk evolution.

Nevertheless, the presence of fragments in the early disk evolution and the lack of 
them in the late evolution implies the existence 
of efficient mechanisms leading to their loss and/or destruction. Among such mechanisms 
are inward migration of fragments onto the protostar 
caused by the gravitational exchange of angular momentum with spiral arms and other fragments
\citep{VB2006,VB2010,Machida2011,Tsukamoto2014}, dispersal of fragments due to tidal torques 
\citep{Boley2010,Zhu2012}, and ejection of fragments from the disk into the intracluster medium 
due to multi-fragment gravitational interaction  
\citep{BV2012}. In our case, all three mechanisms were found to be at work. 
In particular, the inward migration and infall of fragments onto the star triggers 
mass accretion bursts discussed in more detail in Section~\ref{accrete}, while
the ejection of a fragment with mass on the order 
of $0.1~M_\odot$ at $t\approx 0.25$~Myr results in a notable drop in the net disk mass
and a consequent decrease in the disk fragmentation activity.

The $M_{\rm core}$=1.5~$M_\odot$ model~2 is characterized by a greater initial core mass 
than that of model~1. Nevertheless, the evolution in both models is qualitatively similar --
the disk is vigorously unstable to fragmentation in the early evolution showing multiple
fragments interconnected with dense spiral filaments, but  becomes
considerably smoother after $t=0.5$~Myr. This transformation is caused 
by the same effect in both models -- a fragment is ejected form the disk leading to an 
appreciable drop in the disk mass
and subsequent disk stabilization. The disk size and mass in model~2 at later times are somewhat 
smaller than those in model~1, but this is a mere consequence of a somewhat more massive fragment
being ejected from the disk in model~2 ($\sim 0.25~M_\odot$).

On the other hand, model~3 with a smaller initial core mass $M_{\rm core}$=0.3~$M_{\odot}$ demonstrates
a drastically different evolutionary pattern from that of models~1 and 2 -- the disk exhibits a 
flocculent spiral structure in the early evolution ($t\le 0.1$~Myr) and becomes increasingly 
axisymmetric with time. No fragmentation is evident in the disk (but see Figure~\ref{fig3} below). 
Model~3 owes its special behaviour to the fact that
cores with lower mass (but with similar $\beta$) form lower mass disks. In the case of model~3, the
maximum disk mass is $0.09~M_\odot$ at $t\approx 0.1$~Myr and it gradually drops 
to $0.04~M_\odot$ at $t=1.0$~Myr. According to figure~1 in \citet{Vor2013}, a model with $M_{\rm core}=0.31~M_\odot$,
$\beta$=0.88\%, and the disk mass of 0.09~$M_\odot$ lies very near to the fragmentation 
boundary in the
$\beta-M_{\rm core}$ phase space. This means that model~3 can at best 
produce a couple of isolated fragmentation episodes, which might have been missed in Figure~\ref{fig1}
due to infrequent time sampling.

Model~4 is designed to highlight the disk evolution resulting from the collapse of a cloud core
with low angular momentum. In this model, the ratio of rotational to gravitational energy 
$\beta$ is set to 0.14\%, more than six times smaller than that in model~1, but both models have 
the same value of the initial core mass, $M_{\rm core}$=1.1~$M_\odot$.
As a direct consequence of the low initial angular momentum in the core, the disk radius 
in model~4  is rather small 
($\approx 50$~AU at $t\le 0.1$~Myr and $\la 150$~AU in the later evolution), 
much smaller than in other models. No wonder that there was only one fragment formed at
$t=0.36$~Myr and that one dispersed  after a few orbital period (without reaching the sink cell), 
possibly due to tidal torques from spiral arms. 

\begin{figure}
 \centering
  \resizebox{\hsize}{!}{\includegraphics{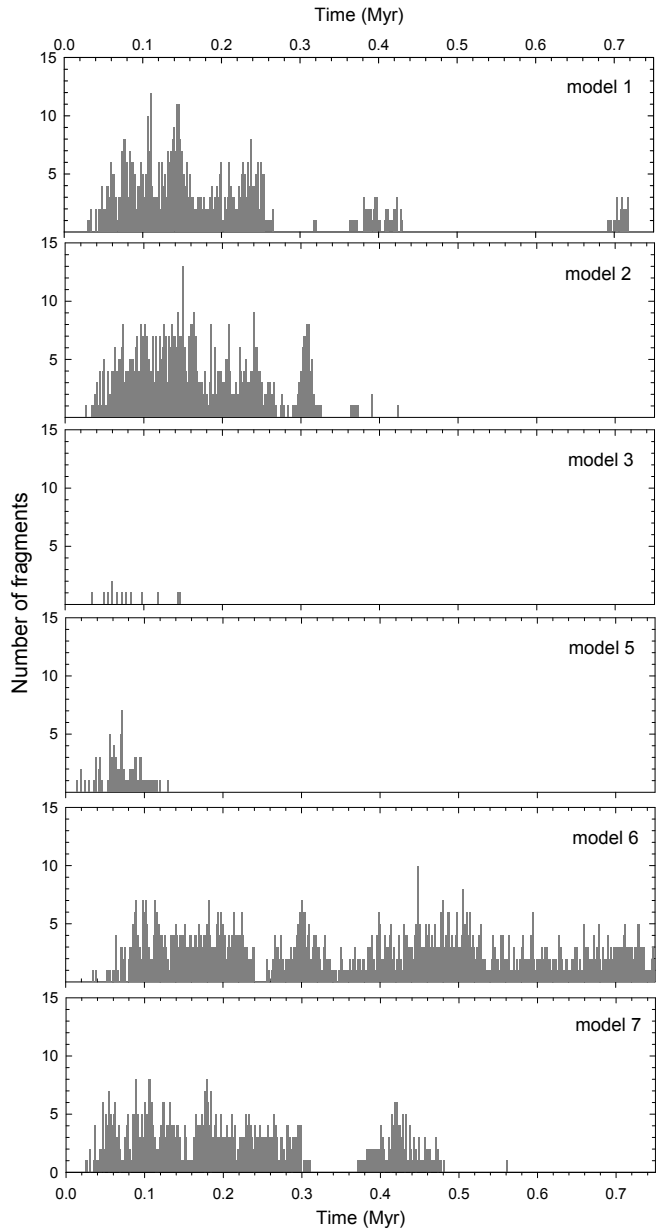}}
  \caption{Number of fragments vs. time in models~1, 2, 3, 5, 6, and 7 (from top to bottom). 
  The number of fragments at a given time instant is calculated using the fragment tracking algorithm
  described in \citet{VZD2013}. An increase in the number of fragments shows recent fragmentation, 
  and a decrease shows recent destruction/accretion of the fragments. The time is counted since the
  formation of the protostar.}
  \label{fig2}
\end{figure}

Model~5 is set to imitate the effect of a warmer  star formation 
environment  with a background radiation temperature of $T_{\rm bg}=25$~K, 2.5 times 
higher than that in model~1. An increase in $T_{\rm bg}$ has notably stabilized the disk against 
fragmentation. There are still occasional fragmentation episodes taking place in model~5 
(see Figure~\ref{fig2} below), but the disk quickly stabilizes even against gravitational 
instability (put aside fragmentation!) and becomes virtually axisymmetric after $t=0.5$~Myr. 

The last but one row in Figure~\ref{fig1} presents the disk evolution in model~6 characterized
by the $\alpha$-parameter equal to 0.03, six times greater  than in model~1. 
The other parameters are identical in both model 6 and 1. A higher efficiency of viscous 
mass and angular momentum transport does not suppress fragmentation, even though the 
disk mass decreases from a maximum value of $0.25~M_\odot$ in model~1 to $0.15~M_\odot$ 
in model~6. Most curiously, model~6 demonstrates a survival of several fragments orbiting the 
host star at wide separation orbits\footnote{We extended the run time to 1.0~Myr and confirm that
the fragments were still present in the disk.}.  This is a rare event, according to numerical simulations
of \citet{VB2010b} and \citet{Vor2013} taking place in one out of ten models with similar characteristics.
We do not think that the fragment survival is caused by a higher value of $\alpha$, since the aforementioned
studies adopted a smaller value of $\alpha=5\times 10^{-3}$. Instead, the present simulations suggest
that the fragment survival can take place for a wide range of $\alpha$, meaning that this is a robust
phenomenon.

One possible reason why model~6 demonstrated the fragment survival, while models~1 and 2 did not, 
is that the latter models experienced fragment ejection, losing $0.1~M_\odot$ 
and $0.25~M_\odot$ of the disk mass, respectively. 
This led to significant weakening of gravitational instability and virtual termination of disk fragmentation
in the later evolution. On the other hand, model~6 lost only $0.02~M_\odot$ via ejection, which 
did not affect appreciably  the strength of gravitational instability and fragmentation. The disk 
in model~6 continued to experience fragmentation even after 0.5~Myr of evolution, which greatly 
increased the odds for fragment survival.
It appears that the gravitational interaction between fragments in the disk is 
intrinsically a {\it chaotic process}, leading in some models to fragment ejection and in others 
to fragment survival.

Finally, the bottom row in Figure~\ref{fig1} presents  the disk evolution in model~7 
characterized by the non-zero mass-to-flux ratio $\mu_{\rm B}=3.33$. Other parameters in model~7 
are identical to those of the non-magnetized model~1. The visual comparison of models 1 and 7 
reveals that the frozen-in magnetic field does not significantly change the
disk propensity to fragment: the fragments are present in the disk during its early evolution. 
In the late evolution (after $t=0.5$~Myr), the disk in the magnetized model seems to be more extended
than its non-magnetized counterpart, but both show no signs of fragmentation.
More accurate numerical simulations with non-ideal magnetohydrodynamical effects, such as 
ambipolar diffusion and magnetic braking, are planned for the near future.

We now analyze the efficiency of disk fragmentation in each model using a much higher time sampling
than in Figure~\ref{fig1}.
Since we do not introduce sink particles to replace fragments in the disk in our Eulerian numerical
code, it is very difficult to track the position and the fate of every fragment during the simulations.
Therefore, we used the fragment detection algorithm described in detail in \citet{VZD2013} to
postprocess our results and calculate the number of fragments present in the disk at a given time. 
We discard fragments that are resolved
by less than 10 grid cells (3 cells in each direction) since their identification on the numerical
grid may be dubious. 
Figure~\ref{fig2} shows the number of fragments $N_{\rm f}$ calculated every 
2000~yr after the formation of the protostar in models~1--7 (from top to bottom). 
Evidently, $N_{\rm f}$ varies significantly with time and from model to model. 
An increase in the number of fragments indicates recent fragmentation, 
and a decrease implies recent destruction/accretion/ejection of the fragments.

The $M_{\rm core}=1.1~M_\odot$ model~1 demonstrates a strong disk fragmentation activity 
in the early evolution. The mean number of
fragments in a time period between 0.03~Myr and 0.26~Myr is four. Episodically, 
the number of fragments may exceed 10 or drop to just a few. This means that the fragment 
formation and destruction mechanisms are constantly at play in the disk. 
At $t\approx0.25$~Myr a massive fragment with some circumfragment material is ejected from the 
disk, reducing the total disk mass and weakening the gravitational instability. Part of the 
ejected material later falls back onto the disk triggering two isolated episodes 
of disk fragmentation at $t\approx 0.4$~Myr and $t\approx0.7$~Myr but none of those fragments survive.
The fragmentation activity in the $M_{\rm core}=1.5~M_\odot$ model~2 is rather similar to that in 
model~1 except that there are no
late-time fragmentation episodes in the disk, probably due to the fact that too much disk 
mass was lost during the ejection episode at $t\approx0.32$~Myr and little fell back onto the disk.
The  $M_{\rm core}=0.3~M_\odot$ model~3 shows just a few isolated disk fragmentation episodes 
with the number of fragments hardly exceeding one at a time.  We do not show model~4 as it 
demonstrated no disk fragmentation. 

The disk fragmentation activity in the $T_{\rm bg}$=25~K model~5 is confined only to the initial 0.1~Myr
of disk evolution. The likely explanation is that disk fragmentation in this model 
is driven by mass infall from the collapsing envelope, which is higher than in other models ($\dot{M}_{\rm
infall} \propto c_{\rm s}^3$). As Figure~\ref{fig6} in Section~\ref{accrete} demonstrates, 
$\dot{M}_{\rm infall}$ is maximal at the time of the protostar formation and quickly drops afterwards.
The disk fragmentation process  in the $\alpha$=0.03 model~6 continues for the whole duration
of the simulation. We saw in Figure~\ref{fig1} that several fragments managed
to survive through almost 1.0~Myr of disk evolution. Figure~\ref{fig3} reveals that the number
of fragments in the late evolution varies between one and just a few, meaning that actually only
one fragment has evolved into a stable companion on a wide orbit. Other fragments form in the 
disk around the companion and in the spiral density waves
connecting the companion with the primary disk rather than in the disk around the central
star. This interesting effect will be studied in more detail in a followup paper.

Finally, the bottom panel in Figure~\ref{fig2} presents the number of bursts vs. time
in the magnetized model~7. Evidently, the frozen-in magnetic field with a mass-to-flux ratio 
of 3.33 does not appreciably change the disk fragmentation activity. The maximum number of fragments
present in the disk at a specific time ($N_{\rm fr}^{\rm max}=8$) is somewhat smaller in model~7 
than in the  corresponding non-magnetized model~1 ($N_{\rm fr}^{\rm max}=12$), but otherwise 
the time behaviour of $N_{\rm fr}$ is similar
in both models.


\subsection{Accretion and infall rates}
\label{accrete}
In this section, we analyze the time behavior of mass accretion rates onto the star ($\dot{M}$) 
and infall rates onto the disk ($\dot{M}_{\rm infall}$) obtained in models~1--6. 
We note that $\dot{M}$ is calculated at the position of the inner sink cell, $r_{\rm cs}=6$~AU,
while $\dot{M}_{\rm infall}$ -- at a distance of 2000~AU.

\begin{figure}
\centering
  \resizebox{\hsize}{!}{\includegraphics{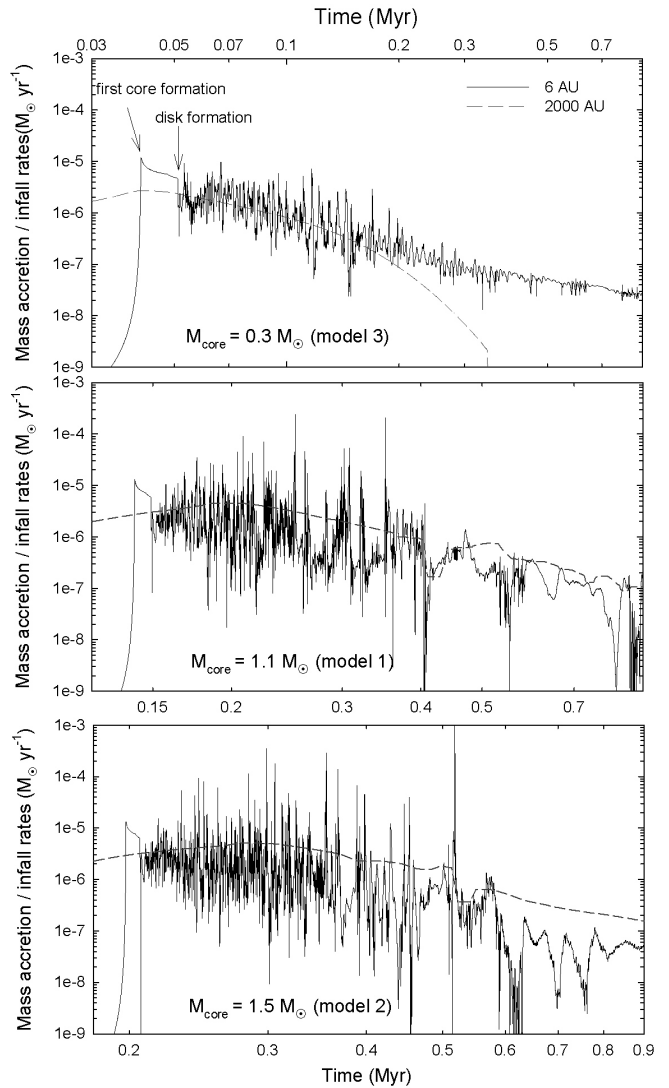}}
  \caption{Mass accretion rates at 6~AU (black solid lines) and envelope infall rates at 
  2000~AU  (red dashed lines) in models~1--3. The arrows mark the formation of the first hydrostatic
  core and the disk in model~1.}
  \label{fig3}
\end{figure}

Figure~\ref{fig3} presents $\dot{M}$ (black solid lines)  and $\dot{M}_{\rm infall}$ (red dashed 
lines) vs. time in models 1--3, which
are characterized by different initial core masses $M_{\rm core}$ as indicated in each panel. 
Other parameters in these models are similar. The time is counted from the beginning of 
numerical simulations (i.e., from the beginning of the core contraction). The formation
of the central protostar in models~1--3 occurs around 0.15~Myr, 0.2~Myr, and 0.05~Myr, 
respectively. Evidently, accretion variability increases with increasing $M_{\rm core}$. 
While the $M_{\rm core}=0.3~M_\odot$ model~3 shows only an order of magnitude variations 
in $\dot{M}$ with just a few stronger fluctuations, the $M_{\rm core}=1.1~M_\odot$
and $M_{\rm core}=1.5~M_\odot$ models 1 and 2 are characterized by highly variable accretion
spanning several orders of magnitude with peak values exceeding in magnitude 
$10^{-4}~M_\odot$~yr$^{-1}$. These accretion bursts are caused by fragments 
spiralling into the central star due to the loss of angular momentum via gravitational 
interaction with spiral arms and other fragments in the disk \citep{VB2005,VB2006,VB2010}.
The animation of this process can be viewed at {\tt http://www.astro.uwo.ca/$\sim$basu/movies.html}. 
We note that in model~2 one accretion burst 
at $\approx0.52$~Myr is especially strong, exceeding in magnitude $10^{-3}~M_\odot$~yr$^{-1}$.
This burst is coeval with the fragment ejection event discussed in the previous section, a pair phenomenon
that causes one fragment to fly into the intracluster medium  and the other  to fall onto the star 
\citep[for more detail see][]{BV2012}.

We now discuss the reason why the time behaviour of mass accretion rates in models with increasing
$M_{\rm core}$ is so different. Figure~\ref{fig1} has already 
given us a hint -- gravitational instability in models~1 and 2 is notably stronger 
than in model~3. This is not surprising considering that more massive cores are supposed to
form more massive disks. To quantify the strength of gravitational instability in the disks 
of our models, we calculated global Fourier amplitudes using the following equation:
\begin{equation}
C_{\rm m} (t) = {1 \over M_{\rm d}} \left| \int_0^{2 \pi} 
\int_{r_{\rm sc}}^{R_{\rm d}} 
\Sigma(r,\phi,t) \, e^{im\phi} r \, dr\,  d\phi \right|,
\label{fourier}
\end{equation}
where $M_{\rm d}$ is the disk mass, $R_{\rm d}$ is the disk's physical 
outer radius, and $m$ is the number of the spiral mode.  
When the disk surface density is axisymmetric, the amplitudes of all modes are
equal to zero. When, say, $C_{\rm m}(t)=0.1$, the perturbation amplitude of 
spiral density waves in the disk is 10\% that of 
the underlying axisymmetric density distribution.

\begin{figure}
\centering
  \resizebox{\hsize}{!}{\includegraphics{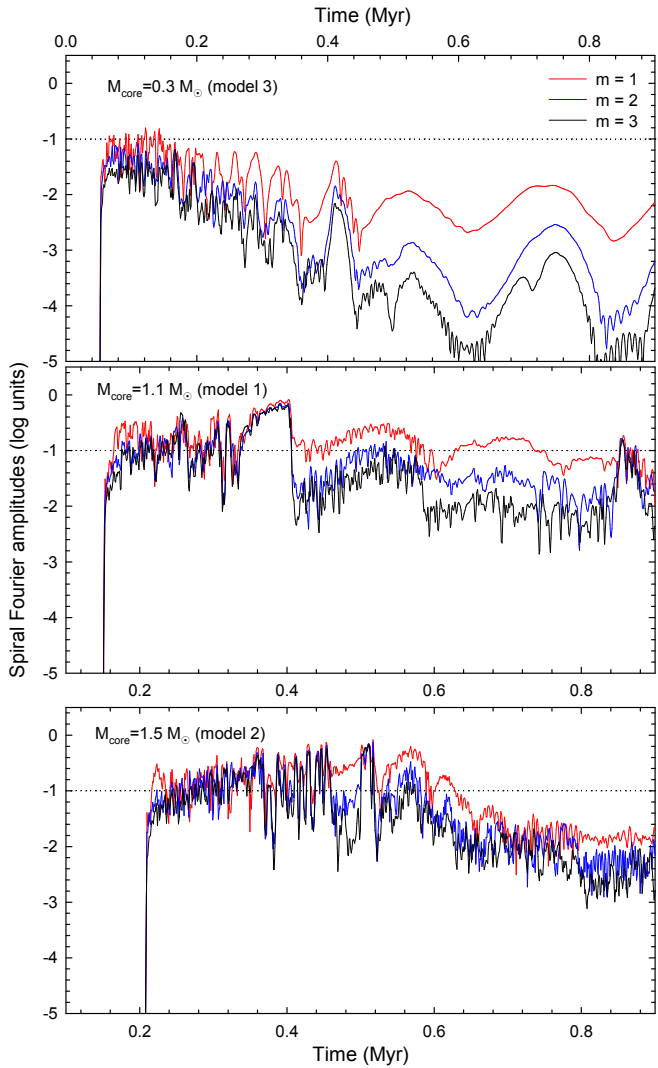}}
  \caption{Global Fourier amplitudes of the first three spiral modes $m$=1--3 in models~1--3. 
  The horizontal dotted lines mark the boundary
  above which spiral waves have perturbation amplitudes greater than 10\% that of 
  the underlying axisymmetric density distribution.}
  \label{fig4}
\end{figure}

The Fourier amplitudes of the first three spiral modes are presented 
in Figure~\ref{fig4} for models 1--3. Evidently, the $M_{\rm core}$=0.3~$M_\odot$ model~3 is 
characterized by the lowest Fourier 
amplitudes -- they hardly exceed 10\% that of the underlying axisymmetric density distribution 
in the early evolution and quickly decline with time. 
The lower-order modes are generally higher in amplitude than the higher-order modes.
On the other hand, the amplitudes of spiral modes in models~1 and 2 are appreciably higher than 
in model~3, reaching values as high as 60\% that of the underlying
axisymmetric distribution ($\log C_{\rm m} \approx -0.2$) by the end of the embedded phase.
In the later evolution, the Fourier amplitudes gradually decline with time, reflecting the overall
disk stabilization due to continuing loss of disk material via accretion onto the star.

When comparing Figures~\ref{fig3} and \ref{fig4}, we can notice a general 
correlation between Fourier amplitudes and variability in the mass accretion rates. 
The low-$M_{\rm core}$ model~3 is characterized by both low Fourier amplitudes
and low accretion variability. As $C_{\rm m}$ declines with time, the variability in $\dot{M}$ 
diminishes. For $C_{\rm m}\la 0.01$, the non-axisymmetric density perturbation in the 
form of spiral density waves is only 1\% that of the underlying axisymmetric distribution, 
implying that the mass transport is now mostly controlled by viscous
torques \citep{VB2009}. The latter drive the disk toward an axisymmetric state as evident 
in Figure~\ref{fig1} and the corresponding accretion rates show only low-amplitude flickering.
On the other hand, Fourier amplitudes in the higher-$M_{\rm c}$ models~1 and 2 are 
greater than in model~3 and the variability in the corresponding accretion rates is notably
higher. Fourier amplitudes comparable to or greater than 10\% imply the presence of 
strong non-axysimmetry in the disk in the form of both spiral waves and fragments.  
 It is important to note that Fourier amplitudes are highly variable\footnote{
In fact, the time variability in $C_{m}$ is even higher but we had to smooth Fourier amplitudes 
somewhat to make them discernible in the figure.} due to the dynamical interaction 
of spiral density waves and fragments in the disk. A combination of two effects: 
that of the nonlinear 
interaction between different spiral modes and fragments in the disk and that of the 
fragments spiralling down onto the star produces variable accretion with episodic  bursts.

\begin{figure}
  \resizebox{\hsize}{!}{\includegraphics{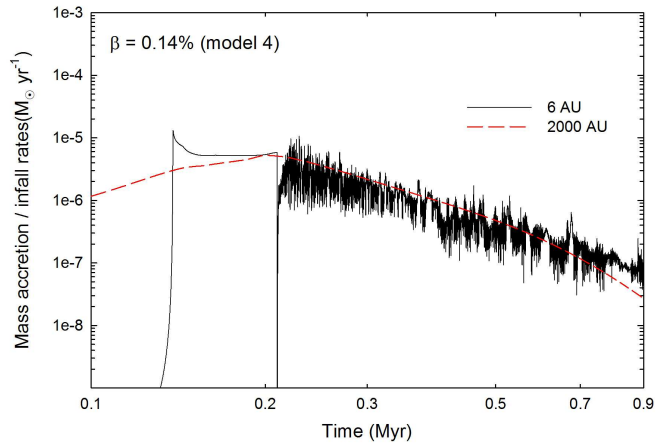}}
  \caption{Mass accretion rates at 6~AU (black solid lines) and envelope infall rates at 
  2000~AU  (red dashed lines) in models~4.}
  \label{fig5}
\end{figure}

We now proceed with analyzing the effect of different values of $\beta$,  $T_{\rm bg}$, 
$\alpha$-parameter  on the variability of $\dot{M}$. 
Figure~\ref{fig5} presents the mass accretion and infall rates in
model~4 characterized by $\beta=0.14\%$, a factor of six lower than in model~1. 
The other parameters in both models are identical. Evidently, the mass accretion
rate in the low-$\beta$ model~4  is characterized by low-amplitude flickering and complete 
absence of bursts, while the high-$\beta$ model~1 shows 
strong accretion variability with multiple bursts, some exceeding in magnitude 
$10^{-4}~M_\odot$~yr$^{-1}$. 
As was demonstrated in \citet{Vor2010}, pre-stellar cores with higher 
$\beta$ tend to form more massive and extended disks than cores with lower $\beta$, which can be understood
on the basis of simple centrifugal radius arguments. 
Massive and extended disks are more gravitationally unstable
and prone to fragmentation than light and compact ones, explaining the aforementioned tendency
in $\dot{M}$.

\begin{figure}
\centering
  \resizebox{\hsize}{!}{\includegraphics{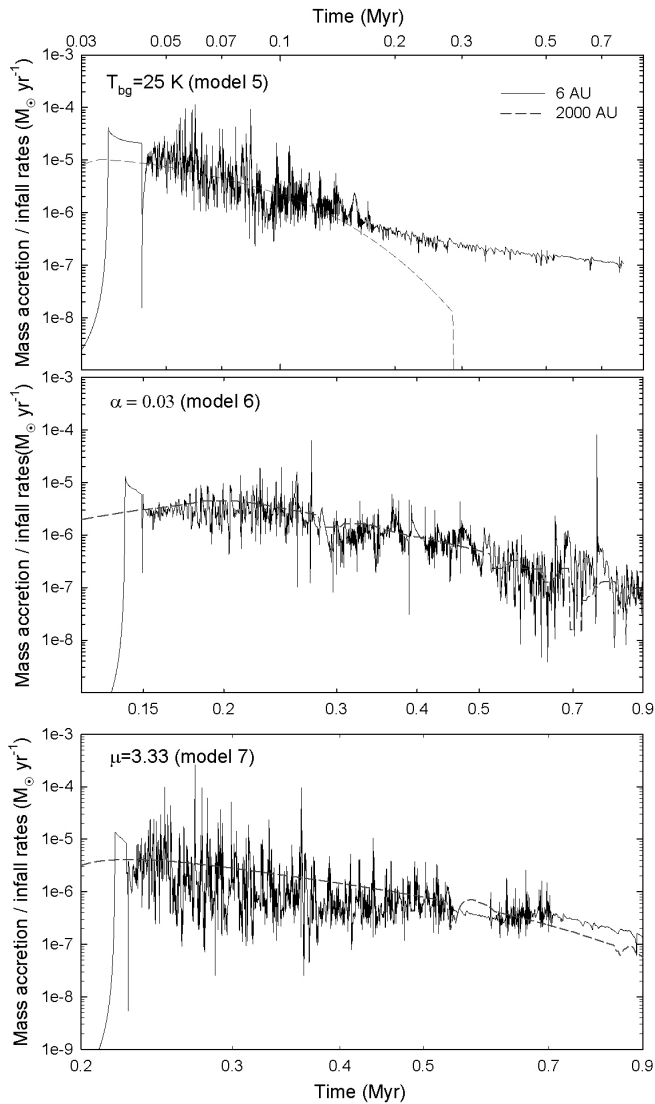}}
  \caption{Mass accretion rates at 6~AU ($\dot{M}$, black solid lines) and envelope infall rates 
  at 2000~AU  ($\dot{M}_{\rm infall}$, red dashed lines) in models~5, 6, and 7 (from top to bottom).}
  \label{fig6}
\end{figure}

As a next step, we describe the effect of a higher background temperature $T_{\rm bg}$ on the 
time behaviour of mass accretion rates, thus mimicking a higher 
heating rate coming from the external environment. The top panel in Figure~\ref{fig6} presents $\dot{M}$
and $\dot{M}_{\rm infall}$ vs. time in model~5 characterized by $T_{\rm bg}=25$~K, 
which is 2.5 times higher than the corresponding value in model~1.  The other parameters are identical
in both models.  The increase in $T_{\rm bg}$ reduces notably the accretion variability. 
This effect is explained by the fact that a higher background temperature raises
the overall disk temperature and reduces the strength of disk gravitational instability.
Nevertheless, the $T_{\rm bg}$=25~K model~5 exhibits several bursts with
$\dot{M}\approx 10^{-4}~M_\odot$~yr$^{-1}$. 

The middle panel in Figure~\ref{fig6} presents $\dot{M}$ and $\dot{M}_{\rm infall}$ 
vs. time in model~6, the parameters of which are similar to those of model~1 except that 
the $\alpha$-parameter is set now to 0.03, a factor of six higher than in model~1.
Evidently, the increase in $\alpha$ acts to reduce the variability in $\dot{M}$, which now 
features only order-of-magnitude variations. Nevertheless, there are two well-pronounced
accretion bursts with $\dot{M}$ approaching $10^{-4}~M_\odot$~yr$^{-1}$. 
The notable reduction in the burst frequency in model~6 as compared to 
the low-$\alpha$ model~1 can be attributed to increased viscous mass transport through the disk, 
which reduces both  the disk mass and the strength of gravitational instability \citep{VB2010}. 
Viscosity also tends to smooth out local non-axisymmetric density enhancements \citep{VB2009}, 
thus reducing the accretion variability caused by spiral density waves.

The last model in this study takes into account the effect of frozen-in magnetic fields.
The bottom panel in Figure~\ref{fig6} presents the mass accretion and infall rates in model~7 which
has the same parameters as the non-magnetic model~1 except that the mass-to-flux ratio
is set to $\mu_{\rm B}=3.33$. Evidently, the magnetized model~7 exhibits several strong 
accretion bursts with a rate $\ge 10^{-4}~M_\odot$~yr$^{-1}$, indicating that the frozen-in
magnetic field does not suppress the burst phenomenon.

Finally, we briefly discuss the time behaviour of infall rates in our models  
and a possible link between $\dot{M}_{\rm infall}$ and
the burst phenomenon. Figures~\ref{fig3}, \ref{fig5}, and \ref{fig6}
demonstrate that  $\dot{M}_{\rm infall}$ steadily increases from 
$\approx10^{-6}~M_\odot$~yr$^{-1}$ to $(5-10)\times10^{-6}~M_\odot$~yr$^{-1}$ during the early
evolution and then gradually declines
below $10^{-7}~M_\odot$~yr$^{-1}$, reflecting the overall depletion of mass in 
the parental cores. Order-of-magnitude variations of $\dot{M}_{\rm infall}$ in the late
evolution of models~1, 2, and 6 are caused by the disturbing influence of fragments that were 
ejected from the disk. A visual inspection of $\dot{M}$ and 
$\dot{M}_{\rm infall}$ in Figures~\ref{fig3}, \ref{fig5}, and \ref{fig6} reveals that 
the amplitude of variations in the accretion rate generally correlates with the infall rate, 
though with some notable exceptions. More specifically, for $10^{-6}~M_\odot~\mathrm{yr}^{-1} \le 
\dot{M}_{\rm infall} \le 10^{-5}~M_\odot~\mathrm{yr}^{-1}$ both the large-scale 
accretion variability and the bursts are  present. For  
$10^{-7}~M_\odot~\mathrm{yr}^{-1} \le \dot{M}_{\rm infall} \le 10^{-6}~M_\odot~\mathrm{yr}^{-1}$
some variability in accretion rates is still present but the bursts are mostly gone, except for model~6
showing one energetic burst in the late evolution.
For $\dot{M}_{\rm infall} \le 10^{-7}~M_\odot~\mathrm{yr}^{-1}$ the accretion variability diminishes
and no bursts are seen. The character of accretion for various infall rates is summarized in Table~\ref{table2}.

\begin{table}
\center
\caption{Infall rates and the character of  accretion}
\label{table2}
\begin{tabular}{ccc}
\hline\hline
$\dot{M}_{\rm infall}$ & Variability & Bursts \\
($M_\odot$~yr$^{-1}$) &  (orders of mag.) &    \\
\hline
$10^{-5}$--$10^{-6}$ & 2--3 & multiple \\ 
$10^{-6}$--$10^{-7}$  & $\sim 1$ & occasional \\ 
$<10^{-7}$  & $<1$ & no \\ 
\hline
\end{tabular}
\end{table}

Model~4, however, stands apart and shows little accretion variability
and no bursts even though the infall rates are greater than $10^{-6}~M_\odot$~yr$^{-1}$ 
during the early evolution. Due to its low angular momentum, the $\beta=0.14\%$ model~4 starts forming
the disk considerable later than other models.
As a result, the disk mass and radius are not sufficiently large for gravitational fragmentation
to take place (see also Fig~\ref{fig1}). High infall rates are therefore a necessary but not 
sufficient condition for the development of the burst phenomenon.

To summarize this section, our numerical simulations with the updated code  confirmed the findings
reported earlier in  \citet{VB2010}, namely that an increase in the initial core mass 
$M_{\rm core}$ and/or the ratio of rotational to gravitational energy $\beta$ acts to increase
the amplitude of accretion variability and the number of accretion bursts.
An increase in the $\alpha$-parameter and $T_{\rm bg}$ is found to have the opposite effect --
the accretion variability and bursts diminish but do not cease to exist, at least for reasonable
values of $\alpha$ and $T_{\rm bg}$. This means that accretion variability
and bursts are a robust phenomenon, weakly sensitive to (modest) variations in, e.g., dust opacities,
turbulent viscosity, and stellar radiation.

\begin{figure}
  \resizebox{\hsize}{!}{\includegraphics{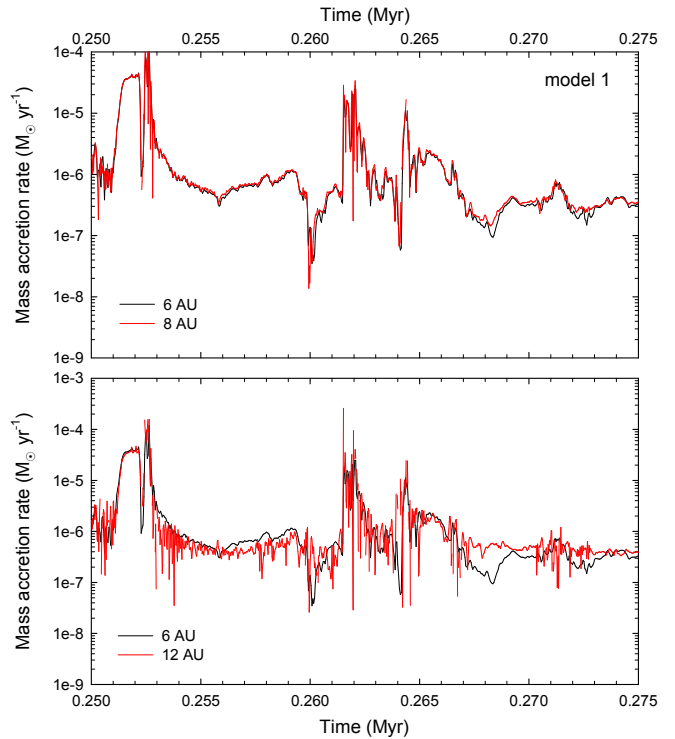}}
  \caption{Comparison of the mass accretion rates at 6~AU and 8~AU (top panel) and 6~AU and 12~AU (bottom
  panel) during a short period of evolution in model~1.}
  \label{fig7}
\end{figure}

\subsection{The effect of the inner boundary condition}
In our models, the mass accretion rate $\dot{M}$ is calculated at the position of the
inner sink cell, $r_{\rm sc}=6$~AU. Two questions arise in this context: how much
$\dot{M}$ is sensitive both to the choice of $r_{\rm sc}$ and to the imposed 
free outflow boundary condition through the sink cell. 
We cannot compute the disk dynamics at sub-AU scales because of a strict Courant time step limitation
imposed on our explicit Eulerian code. However, we performed several test runs with 
$r_{\rm sc}$ varied by a factor of 2 and found little effect on the 
accretion burst phenomenon.

This result can be easily understood by the following reason.
Fragments form in the disk at a radial distance of at least several tens AU from the sink cell
where conditions become favourable for gravitational fragmentation \citep[e.g.][]{Meru2012}. 
At such  distances, the influence of the inner boundary on the disk fragmentation process is
negligible. Furthermore, the typical size of fragments approaching 
the inner sink cell is comparable to the size of the sink cell
itself. Therefore, varying $r_{\rm sc}$ by a factor of  2 does not make much difference --
the fragment will sooner or later pass through the sink cell. 
On the other hand, as fragments approach the star, they must be inevitably
stretched out due to tidal torques. How much of the fragment material finally reaches
the star is an open question and requires a focused investigation \citep[see e.g.][]{Cha2011,Nayakshin2012}.
The effect of a sudden mass deposition onto the inner disk ($\le 10$~AU), as if by infall of a fragment
migrating through the disk onto the star, has recently been investigated by \citet{Ohtani2014}. 
It was found that such an  event can lead to the FU-Orionis-like eruption 
due to triggering of the magneto-rotational instability (MRI) at sub-AU scales. This means that 
either directly, by deposition of material onto the star, or indirectly, by triggering the MRI in the
inner disk, the inward migration of fragments will likely produce luminosity outbursts.

To evaluate the possible influence of the free outflow boundary condition on the 
mass accretion rate through the sink cell, we calculated the mass transport rates at a few AU 
away from the sink cell. Figure~\ref{fig7} presents the mass accretion rates through the sink 
cell at 6~AU 
($\dot{M}$, black lines), as well as the mass transport rates through the disk at 
8~AU ($\dot{M}_8$, red line) and 12~AU ($\dot{M}_{12}$, red lines). 
A narrow time interval of $2\times 10^4$~yr in model~1 is chosen to focus on a time
period featuring both the bursts and quiescent accretion. Evidently, $\dot{M}_8$ is very 
similar to $\dot{M}$ 
except for a few instances when notable deviations are visible. On the other hand, $\dot{M}_{12}$
demonstrates larger deviations from $\dot{M}$, but retains the main qualitative features 
of $\dot{M}$ such as accretion bursts. 
Both $\dot{M}_8$ and $\dot{M}_{12}$ exhibit more variability than $\dot{M}$, most likely due to the
fact that the inner boundary allows for matter to flow into the sink cell but not out of it, thus
somewhat artificially damping the time variations.


\section{Characteristics of luminosity bursts}
\label{bursts}
In this section, we analyze the characteristics of luminosity bursts obtained in our models and compared
them with the available statistics on FU-Orionis-type eruptions (FUors) taken from the
recent review paper by \citet{Audard2014}. 
In order to distinguish the bursts from regular (order-of-magnitude) variability in our models
we have to make several assumptions.
First, we assume that the total luminosity $L_\ast$ during the burst
should be comparable to that of FUors. The latter are usually characterized
by an increase in brightness by at least a factor of 3--4 (in stellar magnitudes) as compared to 
the pre-burst, quiescent phase. Therefore, we stipulate that the luminosity 
increase during the burst should be at least 16 times ($\sim 3$~mag) that of the pre-burst phase. 

Calculating the luminosity in the pre-burst phase turned out to be not an easy task due to 
a highly changeable nature of accretion.
We do this by defining the so-called background luminosity $L_{\rm bg}$, 
which comprises the stellar photospheric luminosity $L_{\rm \ast,ph}$ and the mean accretion luminosity
$\langle L_{\rm \ast,accr} \rangle$. The former is provided by a stellar evolution code (see
Section~\ref{model}), while the latter is found as 
\begin{equation}
\langle L_{\rm \ast,accr} \rangle = {G M_\ast \langle \dot{M} \rangle \over 2 R_\ast},
\end{equation}
where $\langle \dot{M} \rangle$ is the mean accretion rate
calculated using a running average of the instantaneous accretion rates $\dot{M}$
over a time period of $10^4$~yr. When doing the average, we filtered out values that are
greater than $5\times 10^{-6}~M_{\odot}$~yr$^{-1}$ by the 
reason that they may already represent a burst in its rising or fading phase.

\begin{figure*}
   \includegraphics[width=8.5cm]{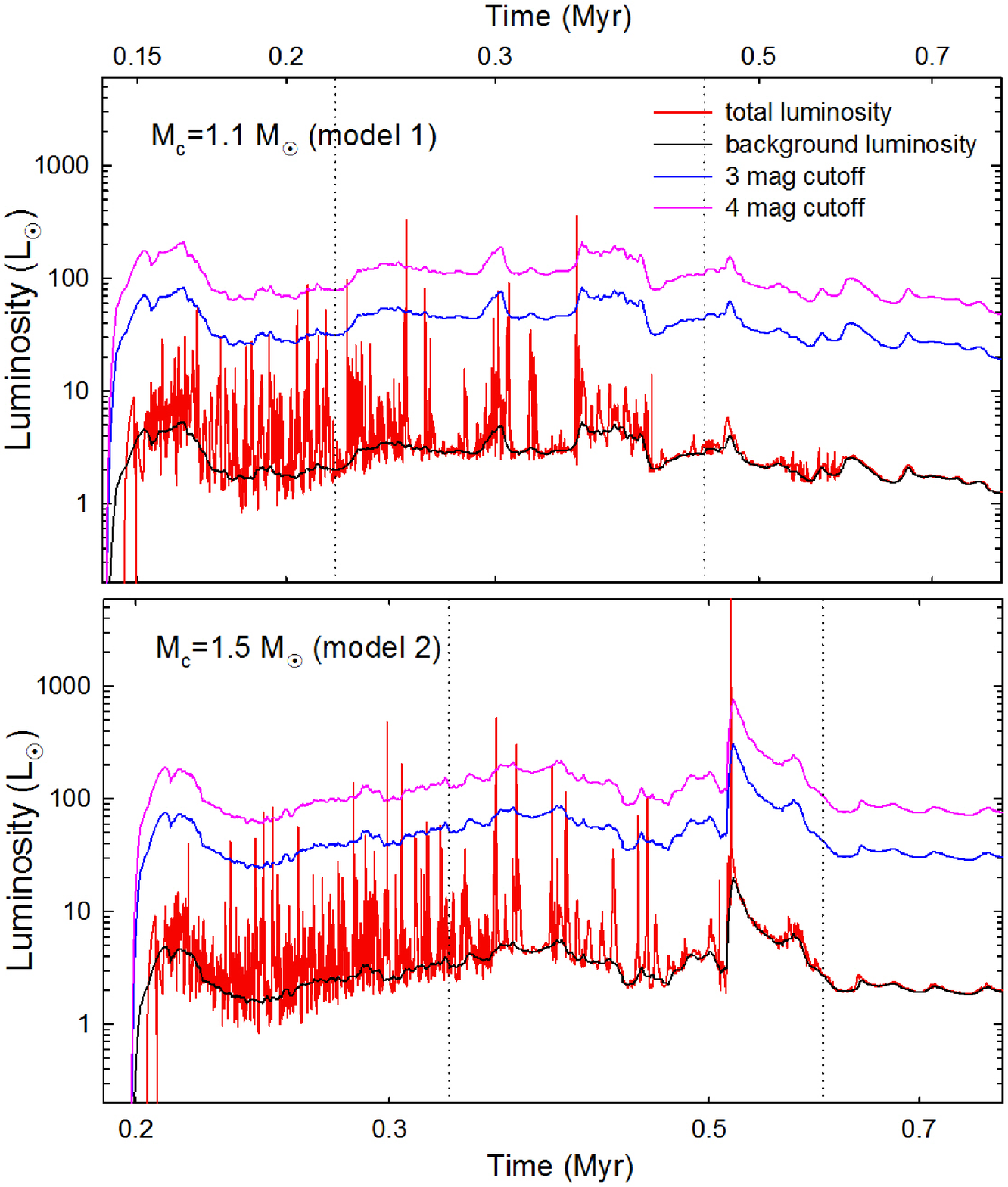}
   \includegraphics[width=8.5cm]{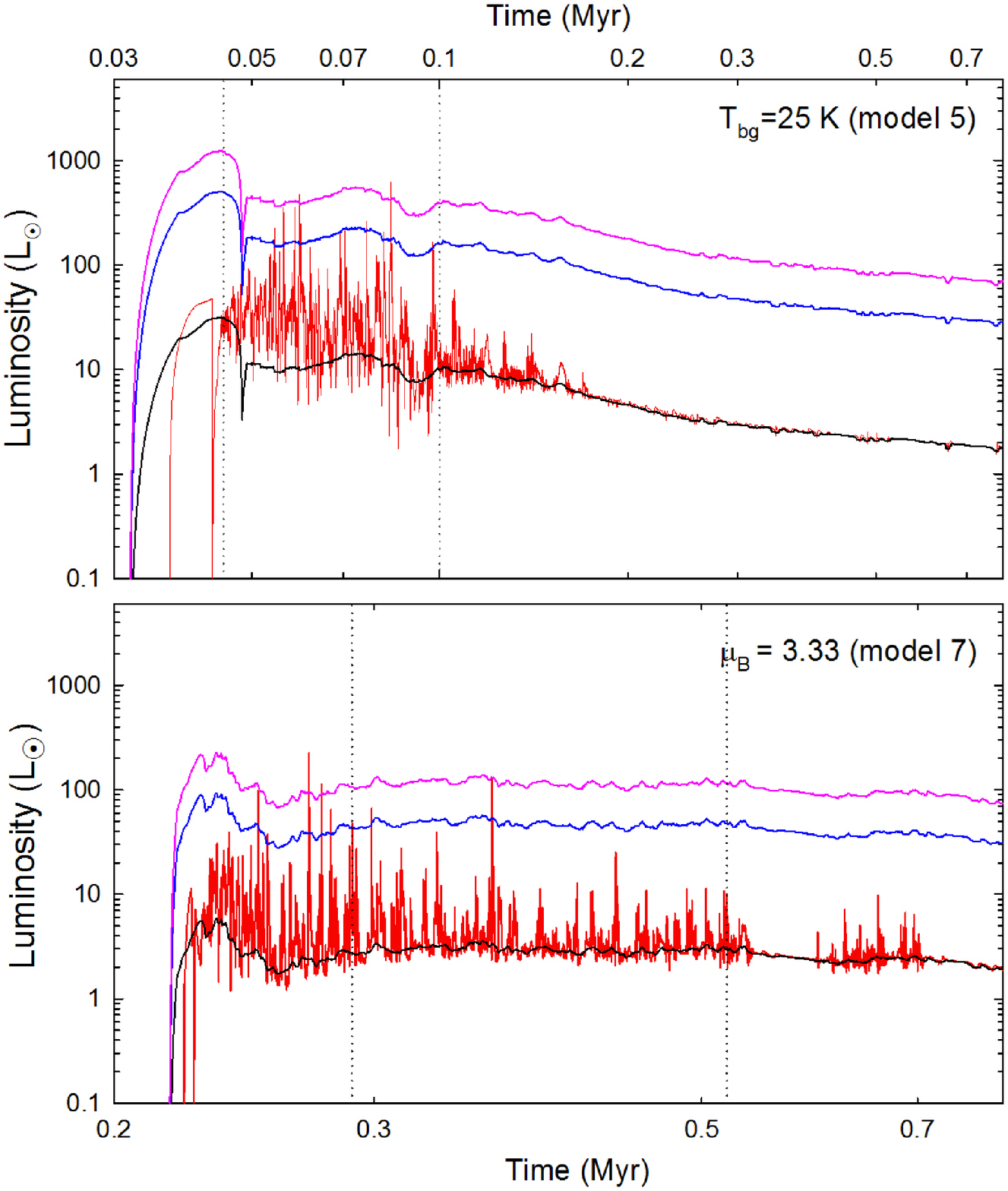}
  \caption{Red lines: total (accretion plus photospheric) luminosity  vs. time in model 1 (top left),
  model~2 (bottom-left), model~5 (top-right), and model~7 (bottom-right).
  The black lines provide the background luminosity comprising the photospheric luminosity plus accretion
  luminosity arising from accretion with a rate $\le 5\times 10^{-6}~M_\odot$~yr$^{-1}$.
  The blue and pink lines mark the 3-magnitude and 4-magnitude cutoffs above which a surge in luminosity
  is considered to be an FU-Orionis-type outburst. The vertical dotted lines mark the Class 0/I boundary
  (left lines) and Class I/II boundary (right lines). See the text for more details.}
  \label{fig8}  
\end{figure*}

The red and black lines in Figure~\ref{fig8} present the total luminosity $L_\ast$ and background luminosity
$L_{\rm bg}$ in models~1, 2, 5, and 7. We left out models that showed too few bursts to be statistically
meaningful. In general, $L_\ast$ is highly variable in the early evolution, reflecting
the corresponding variations in the mass accretion rate. On the other hand, $L_{\rm bg}$ shows much
less variability and describes well the minimal luminosity in each model.
The blue lines mark the values that are 16 times greater than the background 
luminosity at a given time, representing therefore the 3-magnitude cutoff above which a surge in 
luminosity may be classified as a FUor. In principle, these relatively modest
bursts can be confused with the so-called EXors named after its prototype EX Lupi 
\citep[see e.g.][]{Audard2014}. Therefore, with the blue line we also plot the 4-magnitude cutoff
(39 times greater than $L_{\rm bg}$) in order to analyze the statistics of 
more energetic bursts, which are more likely to represent bona fide FUors.

Evidently,  models~1 and 2 are characterized by the largest number of strong FUor-type bursts, 
amounting to 10 and more per model\footnote{Some of the bursts are closely packed and cannot be resolved
in the figure (see Section~\ref{cluster}).}. 
The higher-$T_{\rm bg}$ model 5 has only a few bursts above the 4-magnitude cutoff. At the same
time, model~7 demonstrates several strong bursts despite the presence of frozen-in
magnetic field with a mass-to-flux ratio $\mu_{\rm B}=3.33$, indicating that 
FUors can occur in magnetized  disks as well.

\begin{figure}
  \resizebox{\hsize}{!}{\includegraphics{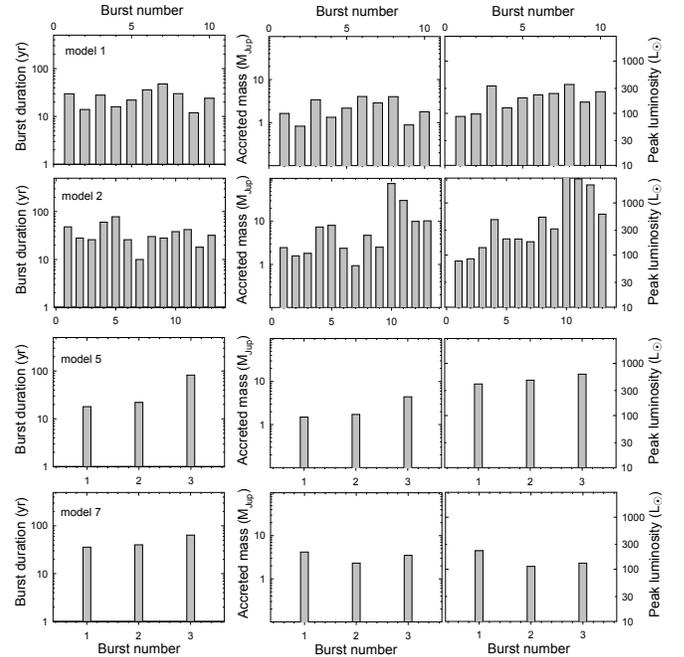}}
  \caption{Burst characteristics in model~1 (top row), 2 (middle row) and 5 (top row). 
  Columns (from left to right) present the duration of the bursts $t_{\rm bst}$ (in yr), 
  the accreted mass during the  bursts $M_{\rm accr}$ (in $M_{\rm Jup}$), and the peak luminosity
  during the bursts $L_{\rm bst}^{\rm max}$ (in $L_\odot$).   }
  \label{fig9}
\end{figure}

In the following text, we analyze the main characteristics of the luminosity bursts obtained 
in our models in order to compare our predictions with observations.
Figure~\ref{fig9} presents the duration of the burst $t_{\rm bst}$ (left column), 
the accreted mass during the burst $M_{\rm accr}$ (middle column), and the peak luminosity
during the burst $L_{\rm bst}^{\rm max}$ (right column) in models~1, 2, 5 and 7 (from top row to bottom
one). Only bursts with a 4-mag cutoff are shown. The $x$-axis shows 
the ordinary number of the burst arranged along the line of the burst occurrence.
Evidently, the burst duration stays mostly in the 10--100~yr limit with 
little dependence either on time or particular model. These values are in good agreement with the 
measured or inferred duration of FUors \citep[see Table \ref{table2} below and 
table~1 in][]{Audard2014}. The accreted mass during the bursts ranges
from 1.0 to 75 Jupiter masses, covering the full mass range of giant planets and brown dwarfs
and reflecting the mass range of fragments forming in the disk \citep{VZD2013}. 
The peak luminosities of most bursts span a range between $75~L_\odot$ and $600~L_\odot$, with a few
notable exceptions in model~2 reaching values in excess of $3000~L_\odot$. These most luminous (and
closely-packed) bursts occur during the fragment ejection event discussed in Section~\ref{disks} and
are the result of the conservation of angular momentum, causing one fragment to fly out of the disk
and  the other fragment to fall onto the star due catastrophic loss of angular momentum during 
the close encounter. 
The fact that there are three closely-packed bursts instead of just one is explained
by tidal destruction of the infalling fragment (see Section~\ref{cluster}).

The summary of various characteristics of the bursts obtained in our modeling are provided 
in Table~\ref{table3}. More specifically, the second column provides the total number of bursts 
$N_{\rm bst}$ and the number of bursts in the deeply embedded 
Class~0 phase (in parentheses, see Section~\ref{embedded}), the third column 
is the fraction of stellar mass accreted 
during the bursts $M_{\rm bst}^{\rm tot}$, and the fourth column is the fraction
of total disk lifetime spent in the burst phase $t_{\rm bst}^{\rm tot}$. The other columns 
(from 5th to 9th) present the maximum, minimum and mean luminosities of the bursts ($L_{\rm max}$,
$L_{\rm min}$ and $L_{\rm mean}$), the maximum, minimum and mean accretion rates during the bursts 
($\dot{M}_{\rm max}$, $\dot{M}_{\rm min}$
and $\dot{M}_{\rm mean}$), the maximum, minimum and mean burst durations 
($t_{\rm bst}^{\rm max}$, $t_{\rm bst}^{\rm min}$ and $t_{\rm bst}^{\rm mean}$), 
the maximum, minimum and mean duration of the quiescent phase
between the bursts\footnote{We calculated these quantities by making no distinction between isolated
and clustered bursts, see Section~\ref{cluster}.} 
($t_{\rm qst}^{\rm max}$, $t_{\rm qst}^{\rm min}$ and 
$t_{\rm qst}^{\rm mean}$), and the maximum and mean accreted mass during the bursts 
($M_{\rm accr}^{\rm max}$ and $M_{\rm accr}^{\rm mean}$).
The mean values were found by arithmetically averaging over all bursts in a particular model.
The known characteristics of FUors are provided in the bottom line and are taken from table~1 of \citet{Audard2014}.

Our models seem to reproduce the main properties of FUors rather well. When averaged over all models,
the mean luminosity during the burst in our models is $312~L_\odot$, about a factor 
of 1.5 greater than that of FUors. The agreement becomes even better after taking out rare 
outliers such as very energetic bursts in model~2.
The averaged burst duration is 39~yr, again almost a factor of 2 greater
than that of FUors, but this may be simply due to the fact that many FUors are 
still found in the active phase. The duration of the quiescent phase also seems to 
be in agreement with the recently estimated lower limit of 5-10~kyr using mid-infrared photometry  
\citep{Scholz2013}. There is however some disagreement which may be of physical nature.
For instance, our model accretion rates agree well with the mean inferred 
mass accretion rates in FUors, but
fail to explain  low-$\dot{M}$ objects such as HBC~722 with 
$\dot{M}\sim 10^{-6}~M_\odot$~yr$^{-1}$ \citep{Audard2014}. 
Indeed, for the minimum accreted mass of 1.0 Jupiter and
the maximum duration of the burst of 100~yr, the resulting minimum accretion rates in our models
are supposed to be around $10^{-5}~M_\odot$~yr$^{-1}$, implying that FUors like HBC~722 may be
driven by mechanisms other than disk fragmentation.

\begin{table*}
\renewcommand{\arraystretch}{1.2}
\center
\caption{Characteristics of luminosity bursts}
\label{table3}
\rotatebox{00}{
\begin{tabular}{ccccccccc}
\hline\hline
Model & $N_{\rm bst}$ & $M_{\rm bst}^{\rm tot}$ & $t_{\rm bst}^{\rm tot}$ & 
$L_{\rm max}/L_{\rm min}/L_{\rm mean}$
& $\dot{M}_{\rm max}/\dot{M}_{\rm min}/\dot{M}_{\rm mean}$ & $t_{\rm bst}^{\rm max}/t_{\rm bst}^{\rm min}/t_{\rm bst}^{\rm mean}$  & 
$t_{\rm qst}^{\rm max}/t_{\rm qst}^{\rm min}/t_{\rm qst}^{\rm mean}$ & $M_{\rm accr}^{\rm max}/M_{\rm accr}^{\rm mean}$ \\
 &  & (\%) & (\%) & ($L_\odot$) &  ($10^{-4}~M_\odot$~yr$^{-1}$) & (yr) & ($10^4$~yr) &  
 ($M_{\rm Jup}$) \\
\hline
& & & & & 4-mag cutoff  & & & \\
\hline
1 & 10(1) & 3.8 & 0.035 & 357/87/208  & 2.4/0.78/1.4 & 48/12/25 & 10/1.6/4.7 & 4.0/2.3 \\
2 & 13(5) & 18.6 & 0.06 & 3042/77/846  & 20/0.8/5.3 & 78/10/36 &  15/0.36/4.4 & 75/12 \\
5 & 3(0) & 1.3 & 0.02 & 620/403/500  & 1.2/0.92/1.0 & 82/18/41 &  -- & 4.4/2.5 \\
7 & 3(2) & 2.0 & 0.04 & 227/115/157  & 2.6/0.9/1.5 & 64/36/47 &  -- & 4.1/3.3 \\
\hline
& & & & & 3-mag cutoff   & & & \\
\hline
1 & 21(9) & 6.4 & 0.086 & 357/28/110  & 2.4/0.25/0.81 & 120/2/31 & 4.3/0.005/1.0 & 9.0/1.9 \\
2 & 42(15) & 24.7 & 0.24 & 3042/25/300  & 20/0.14/2.1 & 320/2/47 &  6.4/0.002/0.8 & 75/6 \\
5 & 15(0) & 5.2 & 0.09 & 620/163/290  & 1.2/0.21/0.64 & 132/14/50 &  1.4/0.004/0.33 & 5.8/2.4 \\
7 & 12(8) & 4.4 & 0.1 & 227/37/81  & 2.6/0.37/0.79 & 90/6/33 &  6.2/0.002/1.1 & 6.7/1.9 \\
\hline
& & & & &    FUors  & & & \\
& & & & & (observations) & & & \\
\hline
 & 26 & -- & -- & 525/10/200  & 10/0.01/1.9 & 80/4/20 & -- & -- \\
\hline\hline
\end{tabular}
}
\end{table*}

\section{Embedded vs. optically visible bursts}
\label{embedded}

According to \citet{Quanz2007}, FUors can be classified in two categories, depending on whether 
silicate features at 10~$\mu$m are seen in absorption or emission.  FUors with silicate in absorption
are likely still embedded in parental envelopes, whereas FUors with silicates in emission are 
likely more evolved and having (partially) depleted envelopes. Out of 21 observed FUors in the 
Quanz et al. sample, 12 have silicates in absorbtion and 7 in emission, with 2 objects having 
a flat spectrum (making their classification dubious). This simple analysis 
suggests that most FUors are rather young objects, possessing sizeable envelopes.

A similar conclusion can be made using data summarized in the recent review on episodic
accretion in young protostars by \citet{Audard2014}.
Two prominent objects, FU Ori itself and V1515, are often considered as most evolved FUors having 
little-to-no envelope, suggested by weak far-infrared/submillimeter continuum beyond 100~$\mu$m
in their spectral energy distributions\footnote{Even in this case, near-infrared interferometry 
shows that, e.g., V1515 may have an undetected contribution from the envelope \citep{Millan-Gabet2006}.}.
According to table~1 in Audard et al., these two objects have the optical 
extinction $A_V$ lying in the 1.5--3.2 range. On the other hand, young FUors with silicate in 
absorption have a minimum extinction of $A_V=4.2$ \citep{Quanz2007}. 
We therefore set $A_V=4.0$ as a tentative boundary between embedded and optically visible
FUors. In the Audard et al.'s sample, 15 FUors are characterized by 
$A_V>4.0$ and only 6 FUors have $A_V<4.0$. Of course, this simple analysis may somewhat be affected
by inclination: highly inclined FUors may appear embedded, whereas in reality they are not.  
Nevertheless, all above arguments taken together indicate that
many (if not most) FUors are young, embedded objects rather than older, Class II stars.

To classify the bursts in our models, we use the remaining mass in the envelope to define
the boundary between the embedded and optically visible phases. Namely, we assume that
the optically visible Class II begins when less than 10\% of the initial core mass
is left in the envelope. The boundary between the deeply embedded Class 0 phase and 
the partly embedded Class I phase is defined as the time when 50\% of the initial core mass 
is left in the envelope. Our adopted classification scheme is based on physical 
properties of a young stellar object, such as envelope and disk masses 
\citep[e.g.,][]{Robitaille06,Dunham2010,Dunham2014}, rather than on observational signatures,
such as submillimeter luminosities or effective temperatures \citep[e.g.][]{Andre93,Chen95}.
Classifications relying upon physical properties are usually referred in the literature 
as ``stages'', whereas those using observational signatures are called ``classes''.
For simplicity here we use the term ``class'' to refer to both the physical stages
and observational classes. Our adopted definition of physical stages 
was extensively investigated in \citet{Dunham2010}. They found
that there is not always a one-to-one correspondence
between physical stage defined by the envelope mass and observational class 
defined by the submillimeter luminosity or effective temperature due to the effects
of geometry and extinction. In reality the exact point at which to set the class boundaries
is somewhat uncertain, which could shift the duration of the embedded
phase in our models by a factor of order unity in either direction. 
We disentangle the disk and infalling envelope on our numerical grid 
using the algorithm described
in \citet{Vor2011}, which is based on the disk-to-envelope transition density of 
$\Sigma_{\rm crit}=0.5$~g~cm$^{-2}$ and the velocity field in the infalling envelope. 
Varying the value of $\Sigma_{\rm crit}$ by a factor of 5
results in changes of the estimated onset time of different phases by only a few per cent.

The vertical dotted lines in Figure~\ref{fig8} mark the Class 0/I boundary (left lines)
and the Class I/II boundary (right lines). Evidently,
most bursts in our models occur in the partly-embedded Class I phase. For instance, 
model~1 has nine strong bursts (above 4-mag cutoff) taking place in the Class I phase, and only 
one in the Class 0 phase,
while model~2 has only five strong bursts out of 13 occurring in the  Class 0 phase.
The values in parentheses in the second column of Table~\ref{table3} provide
the number of bursts (out of the total number) occurring in the 
Class 0 phase. Out of the total 29 strong bursts, only 8 occurred in the 
deeply embedded Class 0 phase and 21 in the partly-embedded Class I phase.
As similar though less pronounced tendency is found for less energetic bursts (3-mag cutoff), except
for the magnetized model~7, in which most bursts occur in the early Class 0 phase.

Our models notably lack bursts taking place in the optically visible Class II phase. 
This is not surprising since most fragments form in the early evolution, which is 
characterized by most massive and gravitationally unstable disks 
(see Figs.~\ref{fig1} and \ref{fig2}). This phase is also less favourable for 
the survival of fragments owing to strong gravitational and tidal torques which
tend to drive fragments onto the star or destroy them. On the other hand, 
fragments that happen to survive through the embedded phase are more likely
to form stable companions, rather than to migrate onto the star and trigger a burst.

There are however exceptions.  
The middle panel in Figure~\ref{fig6} shows the mass accretion rate in model~6.
The Class II phase in this model starts at $t=0.47$~Myr after the onset 
of gravitational collapse. A strong accretion burst at $t=0.763$~Myr,
corresponding to a luminosity outburst of 373~$L_\odot$,  occurs
well into the optically visible Class II phase. 

\begin{figure}
  \resizebox{\hsize}{!}{\includegraphics{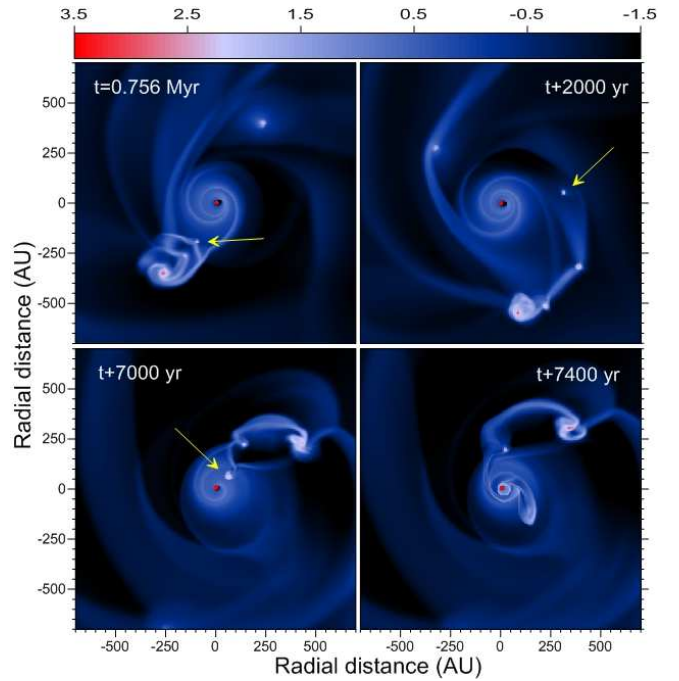}}
  \caption{Gas surface density images (in log g cm$^{-2}$) showing the evolution
  in model~6 before and after the accretion burst at $t=0.763$~Myr. The yellow arrows
  track the position of a fragment forming in the disk of the companion and migrating
  into the central star. The time is counted since the beginning of gravitational collapse. 
  The central star
  is formed at $0.146$~Myr. }
  \label{fig9a}
\end{figure}

Model 6 is unique among other models in the sense that
it reveals the formation of a wide-orbit  companion on a quasi-stable orbit.  
Figure~\ref{fig9a} presents the gas surface density in model~6
in the inner box of $1400\times1400$~AU just before the luminosity outburst
and immediately after it. Evidently, the survived companion possesses
a circumfragment disk, which is sufficiently massive to experience episodic
fragmentation. Indeed, the mass of the companion and its disk
are $57~M_{\rm Jup}$ and $21~M_{\rm Jup}$, making the disk to central object mass ratio 
equal to $\xi\approx 0.38$. Systems with $\xi\ga0.1$ are likely to be unstable to
fragmentation \citep{VB2010}. For comparison, the masses of the central star and its
circumstellar disk at this time are 0.7~$M_\odot$ and $0.022~M_\odot$, and 
the disk to star mass ratio is only $\approx 0.03$, explaining the lack of fragmentation
in the circumstellar disk. The luminosity outburst in model~6 is caused by one of 
the fragments (shown by the yellow arrow) forming in the circumfragment disk and 
falling onto the star owing to a complex
interplay and exchange of angular momentum with other fragments and spiral filaments.

Finally, we note that late bursts in our model are also possible if a fragment is ejected 
from the disk into the intrucluster medium  through the multi-body gravitational interaction 
\citep{BV2012}. The ejection is paired with another fragment losing its angular momentum 
and falling onto the star, producing a strong accretion burst. As table~1 in \citet{BV2012}
demonstrates, some ejection events may occur 0.7--0.8~Myr after the formation of the protostar,
which is usually an optically visible phase. To summarize, most burst events in our model take place
in the partly embedded Class I phase, with a smaller fraction occurring in the deeply embedded phase
and a few bursts in the optically visible Class II phase.

The tendency of luminosity bursts to mainly occur in the embedded phase can be understood by 
analyzing the  updates applied to our numerical hydrodynamics code. 
The use of Semenov opacities (instead of Bell \& Lin's) and stiffer equation 
of state both contribute to make disk fragmentation more difficult owing to an increased
disk temperature. However, the envelope infall onto the disk in the embedded phase increases the 
surface density and brings periodically the disk to the fragmentation boundary. 
Once the envelope begins to dissipate and the infall rate drops, the burst activity subsides,
explaining the scarcity of bursts in the Class II phase. 
The difference with our previous work here is that it now takes pre-stellar cores 
with somewhat higher mass and angular momentum to trigger the burst phenomenon after 
forming a star-disk system.
 
\begin{figure}
  \resizebox{\hsize}{!}{\includegraphics{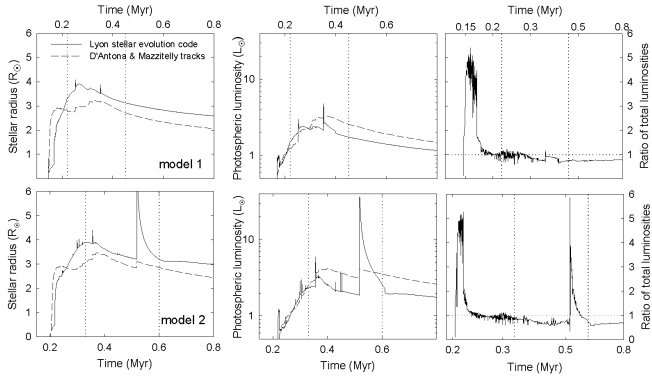}}
  \caption{Time evolution of the stellar radius (left column), photospheric luminosity (middle column)
          and ratio of total luminosities 
          $L_{\rm \ast,Lyon}/L_{\rm \ast,DAM}$ (right column) derived from the Lyon stellar evolution code \citep{Baraffe2010} 
   and the non-accreting stellar evolution tracks of \citet{Dantona94}.
  Shown are results for model~1 (top row) and model~2 (bottom row). The vertical dotted
  lines mark the Class 0/I and Class I/II boundaries. }
  \label{fig9b}
\end{figure}

The use of the Lyon stellar evolution code (instead of precalculated tracks of \citet{Dantona94})
has a more complicated effect. The left and middle columns in Figure~\ref{fig9b} present 
the time evolution of the stellar radius $R_\ast$ and photospheric luminosity $L_{\rm \ast,ph}$
in model~1 (top row) and model~2 (bottom row) derived using the Lyon stellar evolution code 
(solid lines) and D'Antona \& Mazzitelli stellar evolution 
tracks (dashed lines). The right column shows the ratio of total luminosities 
$L_{\rm \ast,Lyon}/L_{\rm \ast,DAM}$ found using the Lyon code and D'Antona \& Mazzitelli tracks
(hereafter, DAM tracks). The vertical dotted lines mark the Class 0/I and Class I/II boundaries.

Evidently, the time evolution of $R_\ast$ is different in the Lyon code and DAM tracks. 
In the early Class 0 phase, the stellar radius in the Lyon code is significantly smaller 
than that of the DAM tracks.   
As a result, the accretion luminosity in the Lyon tracks is considerably greater\footnote{
{Note that when deriving the stellar radii and photospheric luminosities from the DAM 
tracks we use the accretion histories obtained by hydrodynamical simulations coupled with
the Lyon stellar evolution code. That is why the accretion rate and stellar mass are 
identical in both the DAM and Lyon cases}.}. At the same time, the photospheric luminosity in the 
Lyon code and DAM tracks are similar. The net result is that the total luminosity in the Lyon code
is notably greater in the early Class 0 phase, as the right-hand-side column in Figure~\ref{fig9b}
demonstrates, making disk fragmentation more difficult and reducing the burst activity 
in the early Class 0 phase.  

In the late Class~0 phase, the stellar radius in the Lyon code increases owing to
absorption of a fraction of the accretion energy and the situation reversers: 
$R_\ast$ in the Lyon code becomes larger than that in the DAM tracks.
Steep episodic rises in $R_\ast$ seen in model~1 and especially in model~2
at $t\approx 0.55$~Myr are caused by accretion bursts, leading to stellar bloating due to
the absorbed accretion energy.  In the Class I phase, $R_\ast$
in the Lyon code is systematically larger than in the DAM tracks and the photospheric 
and accretion luminosities are smaller (apart from time instances
with strong bursts). The net result is that the total luminosity in the Lyon
code becomes smaller on average that that in the DAM tracks, making disk fragmentation easier.
This effect, along with continuing infall from the envelope, explain why luminosity bursts tend to occur in the Class I phase.

\section{Isolated and clustered luminosity bursts}
\label{cluster}
\begin{figure}
  \resizebox{\hsize}{!}{\includegraphics{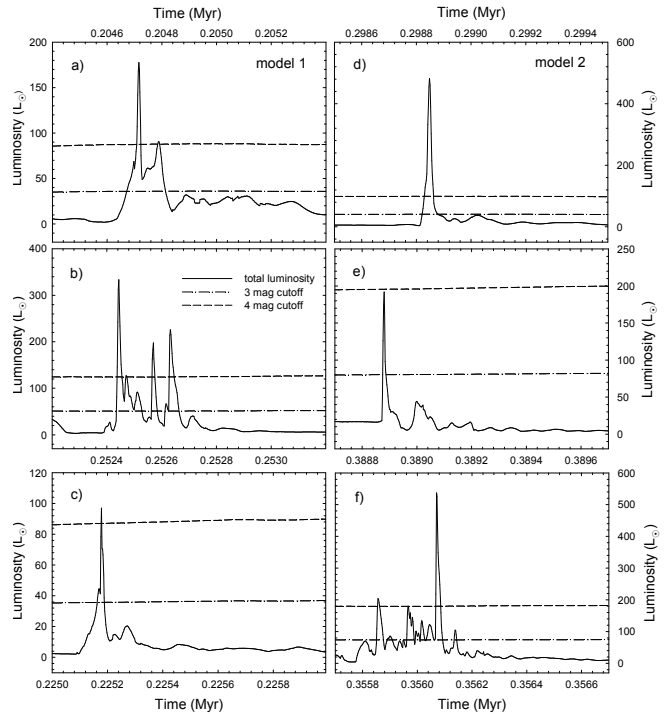}}
  \caption{Zooming in onto  individual bursts in model~1 (left column) and model~2 (right column).
  The solid lines are the total luminosity vs. time, while the dashed and dash-dotted lines 
  are the 4-magnitude and 3-magnitude cutoffs.}
  \label{fig10}
\end{figure}

In this section, we zoom in onto individual luminosity bursts in our models in order
to describe their evolution on short time scales. Figure~\ref{fig10}
presents the total luminosity of individual bursts in model~1 (left column) and model~2 (right column)
during six time periods, each of $10^3$~yr in duration. Also shown with the dashed and 
dash-dotted lines are the 3-magnitude and 4-magnitude cutoffs to help identify the bursts 
(see Section~\ref{bursts}). Two different types of bursts are evident in the figure.
The first type can be described as single, isolated bursts as shown in panels {\bf c}, {\bf d}, 
and {\bf e}. These bursts are caused by compact infalling fragments, which have withstood 
the disruptive effect of tidal torques when approaching the central star and have passed though the
sink cell almost intact. These events are usually characterized
by rather short rise and fall times due to the compact nature of the fragments triggering the bursts.

\begin{figure}
  \resizebox{\hsize}{!}{\includegraphics{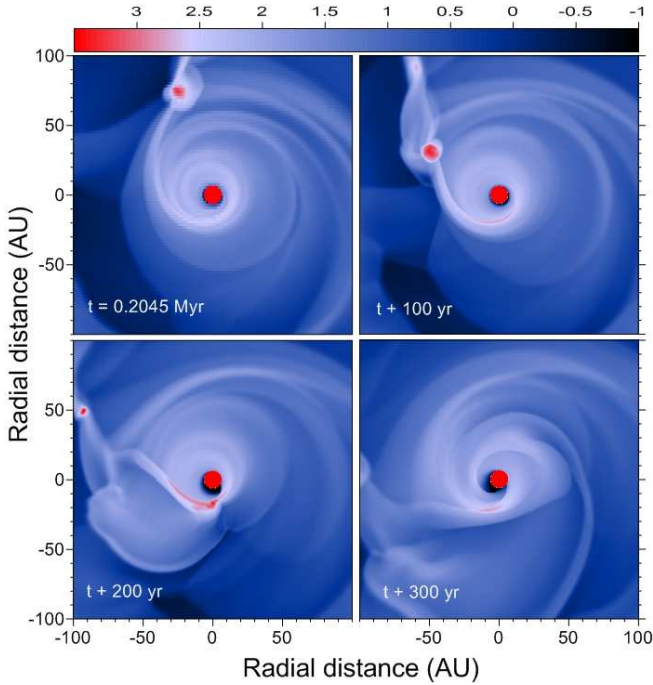}}
  \caption{Zoom-in onto a fragment approaching the central star. Shown is the gas surface density
  (in log g~cm$^{-2}$) during a short time period corresponding to a clustered burst in 
  Figure~\ref{fig10}{\bf a}. The star is marked by the red circle
  in the coordinate center.}
  \label{fig11}
\end{figure}

The other type  are closely-packed bursts, which occur one after another
as shown in panels {\bf a}, {\bf b}, and {\bf f}. As a rule, there is one primary, most 
energetic burst and a few secondary bursts of lesser amplitude. These clustered bursts
are caused by infalling fragments that have started to disintegrate on their approach 
to the star due to the disruptive influence of tidal torques. 
An example of such a phenomenon is illustrated in Figure~\ref{fig11} showing the gas surface density
in model~1 in the inner region during a short time period corresponding to 
Figure~\ref{fig10}{\bf a}. An extended fragment approaching the star is evident in the upper
panels of the figure. At a distance of a few tens AU from the star, 
the fragment loses its roundish shape, stretches into a clumpy filament, and finally
accretes through the sink sell onto the star.  It is interesting to note that another fragment seen
in the bottom-left panel is hurled to a larger distance due to the gravitational exchange 
of angular momentum with the infalling fragment, a phenomenon that can potentially contribute 
to the survival probability of fragments in the disk. 
Depending on the final structure of the filament, two or more bursts of varying amplitude 
can be triggered.
 
The typical times between individual bursts in the clustered-burst event are 30--100~yr, 
too long to be firmly detected on the basis of available FUori observations. 
Nevertheless, the knotty structure seen in many protostellar jets and spacing between the knots 
are consistent with short periods of variability in the mass accretion rates  as could
be expected from clustered bursts.

Another class of young pre-main-sequence stars, EXors (known after its prototype
star EX Lupi), are known for exhibiting repetitive bursts
of lesser magnitude on timescales of several years \citep{Hartmann1996,Audard2014}. 
Until recently, the repetitive nature of
EXor bursts has been considered as a defining characteristic
of such objects. The existence of clustered bursts, if confirmed, could further 
erode the ever shrinking gap between FUors and EXors. In this context, post-burst 
observations of known FUors using (sub)-millimeter interferometers
may search for a clumpy filamentary structure in the inner 20--30~AU 
caused by a disintegrated fragment approaching the star. A successful
detection of such a structure would imply another outburst to come.



\section{The fraction of burst-producing cores}
\label{fraction}
Finally, based on our numerical simulations, we want to estimate the fraction
of star-forming cores that is expected to produce bursts after forming a star-disk system.
Our model with core mass $M_{\rm core} = 0.3\,M_{\odot}$ does not undergo a burst mode 
driven by the formation of large fragments in the disk, however the models we run with 
$M_{\rm core} = 1.1\,M_{\odot}$
and greater mass do show significant bursts with mass accretion rate exceeding $10^{-4}~M_\odot$~yr$^{-1}$.
Figure~6 of \citet{BV2012} shows a correlation 
between fragmentation events, core mass, and $\beta$. Despite limitations of exploration of
parameter space in these models, we can combine the information there with the observational finding
of \citet{Caselli2002} that $\beta$ has a median value of 0.02 and most values in the range $\sim
10^{-4} - 10^{-1}$ to conclude that most cores with $M_{\rm core} \geq 1.0 \,\msun$ will exhibit fragmentation
and bursts and that most cores with $M_{\rm core} \leq 0.5 \,\msun$ will not exhibit bursts. 

To make an estimate of what number fraction of star-forming events these represent, we can further
employ the stellar initial mass function (IMF) of \citet{Chabrier2005} and the idea that the core mass
function is simply the same shape as the IMF but scaled up by a multiplicative factor of about 3 
\citep[see, e.g.,][]{Andre2009}. In this case, and using $0.075\,\msun$ as a minimum stellar mass
\citep{Chabrier2000}, we can identify the number of cores with $M_{\rm core} \geq 1.0 \,\msun$ with the number of stars with mass $m \geq 0.33\,\msun$. Their number fraction can be found easily using the cumulative function of stars introduced by \citet{Basu2014} in their equation~(15): 
\begin{eqnarray} 
F(m) &=& \frac{1}{2} \mathrm{erfc} \left(-\frac{\ln(m)-\mu_0}{\sqrt{2}\sigma_0}\right) \\
& & \hspace{-40pt} -\frac{1}{2}\exp\left(\omega\mu_0+\frac{\omega^2\sigma_0^2}{2}\right)m^{-\omega}
\mathrm{erfc} \left(\frac{\omega\sigma_0}{\sqrt{2}}-\frac{\ln(m)-\mu_0}{\sqrt{2}\sigma_0}\right), 
\label{cumDist}
\end{eqnarray}
where $\mathrm{erfc}(x) = 1- \mathrm{erf}(x)$, $\mathrm{erf}(x)$ is the Error function,
$\omega$ is the power-law index of the high mass tail of the mass distribution function 
($\propto m^{-(1 + \omega)}$), 
and $\mu_0$ and $\sigma_0$ are additional parameters that help define the mean and variance of the distribution.

Their best fit to the Chabrier IMF shows that about 40\% of all stellar objects have 
$m \geq 0.33\,\msun$, hence presumed to originate from a core with mass $M_{\rm core} 
\geq 1.0 \,\msun$. The criterion $M_{\rm core} \leq 0.5\,\msun$ for no bursts will translate 
to $0.075\,\msun \leq m \leq 0.167\,\msun$ in the stellar IMF, and this accounts for some 
32\% of all stellar objects. If 40\% of stellar objects are expected to definitely have bursts 
and 32\% are not, this leaves an intermediate 28\% that may have bursts depending on the level
of rotation in the initial core. Therefore, we can say that the fraction of star-forming cores 
that can be expected to display bursts after forming a star-disk system falls somewhere in 
the range of 40\%-70\%. 

\section{Conclusions}
\label{summary}

We have shown that variable protostellar accretion with episodic bursts is a robust property 
of protostellar collapse. This phenomenon is often present in a collapse environment 
in which a protostellar disk has a self-consistent mass loading from the core envelope. 
An accurate calculation of photospheric and accretion luminosities, as well as
improved disk thermal physics and dust opacities, have refined our numerical hydrodynamics model, 
enabling better characterization of luminosity bursts caused by disk gravitational fragmentation followed
by fragments migrating onto the protostar. 

In agreement with our previous studies, we found that an increase
in the initial core mass and angular momentum favours the burst phenomenon, while higher 
levels of background radiation and magnetic fields can moderate the burst activity.
A minimum mass infall rate onto the disk on the order of a few $\times 10^{-6}~M_\odot$~yr$^{-1}$ 
is required to generate the bursts. The main results can be summarized as follows.
\begin{itemize}
\item A general correlation between the amplitude of time variations in $\dot{M}$ 
and the strength of non-axisymmetric perturbations in the disk, as defined by the global Fourier 
amplitudes, suggests a causal link between accretion variability  and the disk gravitational instability.
In our models, long-term variations in protostellar accretion are caused by the nonlinear 
interaction between different spiral modes in the gravitationally unstable disk, 
while episodic accretion bursts are triggered by fragments migrating onto the star.
\item Most luminosity bursts in our models occur in the partly embedded Class I phase, with a smaller
fraction taking place in the deeply embedded Class 0 phase and a few occasional bursts in the optically
visible Class II phase.   
\item The properties of the bursts  are found to
be in good agreement with those inferred for FU-Orionis-type objects (FUors). For instance,
our models yield the mean luminosity and average burst duration of $312~L_\odot$ and 39~yr, respectively,
only a factor of 1.5--2 higher than in FUors.
\item Depending on the ability of fragments to withstand the tidal torques when approaching 
the star, two types of bursts can occur: the isolated and clustered ones. In the former, 
the fragment is accreted almost intact, producing a well-defined burst, while in the latter 
the fragment is stretched into a knotty filament when approaching the star, 
thus producing a series 
of closely-packed bursts of varying amplitude separated from each other by several decades.
\item Adopting the stellar IMF of \citet{Chabrier2005} and assuming that the core mass
function is the same shape as the IMF but is scaled up by a factor of about 3,  
we estimate that about 40\%--70\%  of the  star-forming cores can be expected to exhibit bursts 
after forming a star-disk system.
\end{itemize} 

In the present work, the mass accretion rates $\dot{M}$ were calculated at the position of 
the inner sink cell at 6~AU. We have shown that the time behavior of $\dot{M}$ is similar 
at 8~AU and 12~AU, thus demonstrating that the accretion variability and the bursts 
are not the artifacts of the adopted inner boundary condition in our numerical simulations. 
Nevertheless, more work is needed to develop 
a self-consistent link between the inner and outer disk regions.

 \section{Acknowledgments}
The authors are thankful to the anonymous referee for providing critical comments that
helped to improve the manuscript.
The authors are thankful to Isabelle Baraffe and Gilles Chabrier 
for providing the stellar evolution code. 
This work is partly supported by the RFBR grant 14-02-00719. The simulations were performed
on the Shared Hierarchical Academic Research Computing Network (SHARCNET),
on the Atlantic Computational Excellence Network (ACEnet), and on the Vienna Scientific
Cluster (VSC-2). SB was supported by a Discovery Grant from the Natural Sciences and
Engineering Research Council (NSERC) of Canada.

\end{document}